\documentclass[sigconf, nonacm, natbib=false]{acmart}

\AtBeginDocument{%
  }

\usepackage{needspace}
\usepackage{booktabs}
\usepackage{algorithm}
\usepackage[noend]{algpseudocode}
\usepackage{xspace}
\usepackage{xcolor}
\usepackage{subcaption}
\usepackage{amsmath}
\usepackage{amsfonts}
\usepackage{bm}
\usepackage{graphicx}
\usepackage{bbm}
\usepackage{esvect}
\usepackage{bbold}
\usepackage{siunitx}
\usepackage{makecell}
\usepackage{paralist}
\usepackage{adjustbox}

\newcommand{\paragraphb}[1]{\vspace{0.03in} \noindent{\bf #1} }

\newcommand{\gandalf}{GANDaLF\xspace}
\newcommand{\ourWF}{NetCLR\xspace}
\newcommand{\netfm}{NetFM\xspace}
\newcommand{\lowbw}{inferior\xspace}
\newcommand{\highbw}{superior\xspace}
\newcommand{\driftcw}{Drift90\xspace}
\newcommand{\driftow}{Drift5000\xspace}
\newcommand{\driftguard}{Drift-guard\xspace}
\newcommand{\drf}{concept drift\xspace}
\newcommand{\Drf}{Concept Drift\xspace}
\newcommand{\semi}{SemiSL\xspace}
\newcommand{\self}{SelfSL\xspace}

\newcommand{\Lowbw}{Inferior\xspace}
\newcommand{\Highbw}{Superior\xspace}

\newcommand{\aug}{NetAugment\xspace}
\newcommand{\flipaug}{FlipAugment\xspace}

\newcommand{\dfaug}{$\text{DF}_\text{augmented-data}$\xspace}
\newcommand{\dfsame}{$\text{DF}_\text{same-data}$\xspace}

\algnewcommand{\LineComment}[1]{\State \(\triangleright\) #1}

\graphicspath{{./images/}}

\RequirePackage[
  datamodel=acmdatamodel,
  style=acmnumeric,
  ]{biblatex}

\setcopyright{none}

\renewcommand\footnotetextcopyrightpermission[1]{} 
\begin{document}
\title{Realistic Website Fingerprinting By Augmenting Network Traces} 

\author{Alireza Bahramali}
\email{abahramali@cs.umass.edu}
\affiliation{%
  \institution{University of Massachusetts Amherst}
  \streetaddress{}
  \city{}
  \state{}
  \country{}
  \postcode{}
}

\author{Ardavan Bozorgi}
\email{abozorgi@cs.umass.edu}
\affiliation{%
  \institution{University of Massachusetts Amherst}
  \streetaddress{}
  \city{}
  \state{}
  \country{}
  \postcode{}
}

\author{Amir Houmansadr}
\email{amir@cs.umass.edu}
\affiliation{%
  \institution{University of Massachusetts Amherst}
  \streetaddress{}
  \city{}
  \state{}
  \country{}
  \postcode{}
}

\begin{abstract}
Website Fingerprinting (WF) is considered a major threat to the anonymity of Tor users (and other anonymity systems). 
While state-of-the-art WF techniques have claimed high attack accuracies, e.g., by leveraging Deep Neural Networks (DNN), several recent works have questioned the practicality of such WF attacks in the real world due to the assumptions made in the design and evaluation of these attacks. 
In this work,
we argue that such impracticality issues are mainly 
due to the attacker's inability in collecting training data in comprehensive network conditions, 
e.g., a WF classifier may be trained only on samples collected on specific high-bandwidth network links but deployed on connections with different network conditions. 
We show that \emph{augmenting network traces} can enhance the performance of WF classifiers in unobserved network conditions.
Specifically, we introduce \aug, an augmentation technique tailored to the specifications of Tor traces.
We instantiate \aug through semi-supervised and self-supervised learning techniques.
Our extensive open-world and close-world experiments demonstrate that under practical evaluation settings, our WF attacks provide superior performances compared to the state-of-the-art; this is due to their use of augmented network traces for training, which allows them to learn the features of target traffic in unobserved settings (e.g., unknown bandwidth, Tor circuits, etc.).   
For instance, with a 5-shot learning in a closed-world scenario, our self-supervised WF attack (named \ourWF) reaches up to $80\%$  accuracy when the traces for evaluation are collected in a setting unobserved by the WF adversary. This is compared to an accuracy of $64.4\%$ achieved by the state-of-the-art Triplet Fingerprinting~\cite{sirinam2019triplet}.
We believe that the promising results of our work can encourage the use of network trace augmentation in other types of network traffic analysis. 

\end{abstract}




\maketitle

\section{Introduction}\label{sec:intro}
Anonymous communication systems hide the identities of the Internet end-points (e.g., websites) visited by Internet users, therefore protecting them against online tracking, surveillance, and censorship.
Tor~\cite{dingledine2004tor} is the most popular anonymous communication system in the wild with several million daily users~\cite{tor-metric}. 
Tor provides anonymity by relaying clients' traffic through cascades of proxies, known as relays.
A major threat to the anonymity provided by Tor and similar anonymity systems is a class of attacks known as \emph{Website Fingerprinting (WF)}~\cite{sirinam2018deep, sirinam2019triplet, cherubin2017website, cherubin2022online, oh2021gandalf, rimmer2017automated, panchenko2011website, cai2012touching, wang2013improved, panchenko2016website, hayes2016k, wang2016realistically, oh2019p1, var-cnn, 184463}. 
WF is performed by a passive adversary who monitors the victim's network traffic, e.g., a malicious ISP or surveillance agency. The adversary compares the victim's observed traffic trace against a set of pre-recorded website traces, to identify the webpage being browsed. 
State-of-the-art (SOTA) WF attacks achieve significantly high accuracies  by leveraging deep neural network (DNN) architectures~\cite{oh2019p1, var-cnn, sirinam2018deep, sirinam2019triplet, cherubin2022online, rimmer2017automated, oh2021gandalf}, e.g., Deep Fingerprinting~\cite{sirinam2018deep} claims $98\%$ accuracy in a closed-world scenario.  

\paragraphb{Critiques of WF Studies:}
Despite the high accuracies claimed by SOTA WF attacks, several recent works~\cite{juarez2014critical, cherubin2022online, critique, wang2016realistically, wang2020high} have questioned the relevance of such attacks in practice due to the (unrealistic) threat models assumed in evaluating such attacks. Notably, the following are the major criticisms: 
\begin{itemize}
    \item \emph{Resilience to concept drift:} Concept drift refers to the phenomenon where the properties that distinguish one website from another can change over time. This can make it more difficult for WF techniques to accurately identify and track individual websites, as the features that were previously used to distinguish them may no longer be reliable. Juarez et al.~\cite{juarez2014critical} show that concept drift causes a significant drop in  WF accuracy.
    
    \item \emph{Network condition variations:} To collect ground truth data for training a WF classifier, researchers usually generate synthetic network traces through automated browsing of a pre-defined set of websites. However, during the deployment of the attack, the WF classifier may encounter traces that are collected in different network conditions (e.g., lower bandwidth). 
    
    \item \emph{Inaccurate user imitation:} Automated browser crawlers such as Selenium~\cite{selenium} are used by researchers to collect ground truth traces.
    The diversity of browser configurations and variations in user behavior such as visiting subpages of a website cannot be replicated by these crawlers.
    
    \item \emph{Requiring large labeled datasets:} Some DNN-based WF techniques require large amounts of labeled data for training to achieve high accuracies. 
\end{itemize}

In response to above criticisms, researchers proposed various heuristic techniques~\cite{cherubin2022online, wang2016realistically, oh2021gandalf, panchenko2016website, rimmer2017automated, wang2020high, pulls2020website} in the design and evaluation of WF attacks. For instance,  Wang and Goldberg~\cite{wang2016realistically} propose to maintain a fresh training set to make the model robust against concept drift. Furthermore, to improve WF accuracy when limited labeled data is available, \gandalf~\cite{oh2021gandalf} and Triplet Fingerprinting (TF)~\cite{sirinam2019triplet} use generative networks and metric learning, respectively. 
Most recently, Cherubin et al.~\cite{cherubin2022online} aim to address the issue of inaccurate user imitation by training a WF classifier on genuine Tor traces collected from exit Tor relays.  

\paragraphb{Enabling WF under realistic settings:}
We argue that the (ad hoc) approaches mentioned above either only partially address the issue, or are impractical themselves. For instance, the SOTA works of  \gandalf~\cite{oh2021gandalf} and Triplet Fingerprinting (TF)~\cite{sirinam2019triplet} only address the issue of data availability, but not concept drift. 
In this work, we argue that the main reason for these issues with WF techniques is the \emph{attacker's inability to collect training network traces in variable network conditions.}
That is, the WF party is either not able to  collect enough (labeled) training  data to represent the diverse and volatile nature of the web traffic over time, or the collected training data only represents a specific threat model, e.g., a specific network condition, a particular Tor circuit, or a specific time frame. 
For instance, the adversary may collect Tor traces in one setting during training but may encounter traces in a completely different setting during the deployment that have not been previously observed. 
One potential solution to mitigate this issue is to collect Tor traces in a variety of network settings and scenarios. 
However, there can be an infinite number of settings and in any practical WF scenario, it is infeasible to collect traces in all possible settings, e.g., to solve the concept drift problem, the attacker needs to re-train the classifier regularly because the contents or even the layout of the websites may change everyday. 

In this work, we aim to alleviate the mentioned issues through \emph{augmentation of network traces}. Augmentation is to modify the existing samples to generate new samples that have the crucial features of the original ones. Our work is inspired by the successful  uses of  data augmentation in various SOTA machine learning architectures, in particular in various emerging semi-supervised and self-supervised applications. 
Our intuition is that augmenting network traces (e.g., of website connections) can help enhance the longitudinal perspective of a WF classifier, by enabling it to obtain network data samples that represent unobserved network conditions or settings. 
However, one can not simply borrow data augmentation techniques designed for classical machine learning tasks like vision. Instead, to get the most out of augmentation, we develop augmentation techniques that are tailored to the specific characteristics of network traffic. 

We demonstrate the impact of data augmentation on boosting the performance of WF attacks under realistic settings. 
We start by evaluating the impact of a naive augmentation approach, i.e., 
 randomly flipping the directions of Tor cells in a Tor trace. 
We then evaluate our network-tailored augmentation mechanism,  \aug, which is  tailored to specific configurations of Tor traffic. 
\aug replicates the modifications that may happen in unobserved WF settings by manipulating bursts of cells in a Tor trace. In our experiments, we show that \aug performs significantly better than a naive (random) augmentation.

\paragraphb{Deploying Network Augmentation:}
Augmented Tor traces can be used in different ways to train a WF classifier.
In this work, we instantiate \aug through semi-supervised learning (\semi) and self-supervised learning (\self). We use \semi and \self to reduce the dependency of the model on collecting large amounts of labeled data.  
After evaluating several deployments of \self and \semi techniques, we propose \ourWF, a WF attack based on \self techniques integrated with \aug. 
\ourWF learns useful representations of Tor traces without any requirement of labeled data as a pre-training phase.
We then fine-tune the pre-trained base model to adjust \ourWF to downstream datasets.
To evaluate \ourWF in a realistic scenario, we perform pre-training and fine-tuning using traces collected in one setting, and perform the attack on traces collected in different settings. 
We split our datasets into two categories:  traces collected in consistent network conditions (\textit{\highbw} traces), and  traces collected in poor and low bandwidth network conditions (\textit{\lowbw} traces).

\paragraphb{Evaluations:}
We perform extensive experiments to evaluate \ourWF in both \textit{closed-world} and \textit{open-world} WF scenarios.
We show that \ourWF outperforms previous WF techniques when the traces for training and deployment are collected in different settings, e.g., in a closed-world scenario and with only 5 labeled samples, \ourWF has $80\%$ accuracy on \lowbw traces while Triplet Fingerprinting (TF), the SOTA low-data WF technique, reaches only $64.4\%$ in an exact same setup.
Furthermore, our results show that \ourWF shows more resilience to \drf compared to previous systems, e.g., when evaluating \ourWF on a dataset with a 5-year gap from the dataset used in pre-training, \ourWF reaches $72\%$ accuracy using 20 labeled samples while TF only reaches $51\%$ accuracy. 
\ourWF is also effective in an open-world scenario, e.g., when using 5 labeled samples, \ourWF reaches $92\%$ precision while Deep Fingerprinting has only $75\%$ precision.

We also evaluate \ourWF against the Blind Adversarial Perturbations (BAP)~\cite{nasr2021defeating} countermeasure technique which is based on adversarial examples and is shown to be effective in defending WF attacks. Based on our results, \ourWF is more robust compared to existing systems when BAP is active, e.g., the accuracy of \ourWF reduces by $4.9\%$ when there are 10 labeled samples while the accuracy of DF decreases by $52.3\%$. 

\paragraphb{Summary of contributions\footnote{Our code and artifacts are available at \url{https://github.com/SPIN-UMass/Realistic-Website-Fingerprinting-By-Augmenting-Network-Traces}}:} 
\begin{compactitem}
    \item We investigate augmentation of  network traffic to alleviate realisticness issues of WF techniques that stem from a lack of longitudinal perspective into network traffic; follow-up studies may extend this to other types of traffic analysis. 
    \item We deploy our network augmentation mechanisms into novel WF techniques that are based on semi-supervised and self-supervised mechanisms. 
    \item We perform extensive experiments in closed-world and open-world settings, demonstrating the superiority of our WF attacks in realistic settings. 
\end{compactitem}

\section{Background}\label{back}

\subsection{Semi-Supervised Learning}\label{sub:semi}

Semi-supervised learning (\semi) is an approach to machine learning that combines a small amount of labeled data with a large amount of unlabeled data during training. 
There is a line of research in \semi aiming at producing an artificial label for unlabeled samples and training the model to predict the artificial label when fed unlabeled samples as input~\cite{lee2013pseudo, mclachlan1975iterative, xie2020self, rosenberg2005semi, scudder1965probability}. These approaches are called pseudo-labeling. 
Similarly, consistency regularization~\cite{bachman2014learning, laine2016temporal, sajjadi2016regularization} obtains an artificial label using the model's predicted distribution after randomly modifying the input or model function. 
FixMatch~\cite{sohn2020fixmatch} is a recent \semi algorithm in computer vision that combines consistency regularization and pseudo-labeling and generates separate \textit{weak} and \textit{strong} augmentations when performing consistency regularization. 
FixMatch produces a pseudo-label based on a weakly-augmented unlabeled sample, which is used as a target label when the model is fed a strongly-augmented version of the same input.
Note that in computer vision, augmentation refers to distorting the pixels of an image to generate new samples with the same label. This includes making small changes to images or using deep learning models to generate new data points. 
In FixMatch, weak augmentation is a standard flip-and-shift augmentation strategy while strong augmentation is based on AutoAugment~\cite{cubuk2019autoaugment}.
We overview the detailed formulation of FixMatch in Appendix~\ref{app:semi}.

\subsection{Self-Supervised Learning}\label{sub:self}

Self-supervised (Unsupervised) learning (\self) tries to learn useful embeddings from unlabeled data. 
Contrastive learning is a subset of unsupervised learning that aims to learn embeddings by enforcing similar elements to be equal and dissimilar elements to be different. 
SimCLR~\cite{chen2020simple} is a recent framework in computer vision for contrastive learning of visual representations. In particular, SimCLR learns representations (embeddings) of unlabeled data by maximizing agreement between differently augmented views of the same data example via a contrastive loss in the latent space. 
Note that SimCLR does not use any labeled data to learn the representations of input samples.
We overview the detailed formulation of SimCLR in Appendix~\ref{app:self}.

\subsection{Website Fingerprinting: An Overview}\label{sub:wf}


Website Fingerprinting (WF) aims at detecting the websites visited over encrypted channels like VPNs, Tor, and other proxies.
\cite{sirinam2018deep, var-cnn, rimmer2017automated, 184463, panchenko2011website, cai2012touching, wang2013improved, panchenko2016website, hayes2016k, wang2016realistically, jansen2018inside, sirinam2019triplet, wang2020high}.
The attack can be performed by a passive adversary who monitors the victim's encrypted network traffic, e.g., a malicious ISP or a surveillance agency.
The adversary attempts to identify the website visited by the victim by observing their encrypted traffic and using various classification techniques.
Website fingerprinting has been widely studied in the
context of Tor traffic analysis~\cite{rimmer2017automated, sirinam2018deep, var-cnn, panchenko2011website, cai2012touching, wang2013improved, panchenko2016website, hayes2016k, jansen2018inside, sirinam2019triplet, wang2020high}.

Various machine learning classifiers have been used for WF, e.g.,  using KNN~\cite{184463}, SVM~\cite{panchenko2016website}, and random forest~\cite{hayes2016k}.
However, SOTA WF algorithms use DNNs to perform website fingerprinting~\cite{sirinam2018deep, rimmer2017automated, var-cnn}.
DNN-based WF attacks demonstrate effective performance in both the closed-world setting, where the user is assumed to only browse websites in a monitored set, and the more realistic open-world setting, where the user might browse any website, whether monitored or not. For instance, Deep Fingerprinting (DF)~\cite{sirinam2018deep} achieves over $98\%$ accuracy in the closed-world setting and over $0.9$ for both precision and recall in the open world by using convolutional neural networks (CNN).
Similarly, Automated Website Fingerprinting (AWF)~\cite{rimmer2017automated} is another algorithm based on CNNs that achieves over $96\%$ accuracy in a closed-world scenario. 
To achieve high accuracies, DF and AWF require large amounts of training data, e.g., DF uses 800 traces per website. Furthermore, both AWF and DF assume the traces for the training and test have the same distribution, while in practice, traces can be collected in different time periods or from different vantage points. 

There are recent studies that perform limited-data WF~\cite{sirinam2019triplet, oh2021gandalf, var-cnn}, i.e.,  Sirinam et al. propose Triplet Fingerprinting (TF)~\cite{sirinam2019triplet} where they examine how an attacker can leverage N-shot learning\textemdash a machine learning technique requiring just a few training samples to identify a given class\textemdash to reduce the amount of data required for training as well as mitigate the adverse effects of dealing with heterogeneous trace distributions. 
TF leverages triplet networks~\cite{schroff2015facenet}, an image classification method in contrastive learning, to train a feature extractor that maps network traces to fixed-length embeddings. TF uses 25 samples per website and 775 websites to train the feature extractor in the pre-training phase. 
The embeddings are then used in a fine-tuning phase to $N$-train a K-Nearest Neighbour (KNN) classifier. 
TF achieves over $92\%$ accuracy using 25 samples per website to train the feature extractor and only 5 samples to train the KNN classifier.  
Generative Adversarial Network for Data-Limited Fingerprinting (GANDaLF)~\cite{oh2021gandalf}, proposed by Oh et al., is another effective WF attack when few training samples are available. In particular, \gandalf uses a Generative Adversarial Network (GAN) to generate a large set of "fake" network traces, helping to train a DNN that distinguishes classes of real training data. In the closed-world setting, \gandalf achieves $87\%$ accuracy using only 20 instances per website (100 sites). 
Online WF~\cite{cherubin2022online} is the most recent WF attack proposed by Cherubin et al. They argue that existing WF techniques lack realistic assumptions making them impractical in real-world scenarios. They show that synthetic traces collected by researchers using automated browsers over entry relays is less diverse than genuine Tor traces.
Therefore, Online WF uses genuine Tor traces collected over an exit relay to perform a more realistic WF attack.  


\section{Problem Statement}\label{problem}


\subsection{Critiques of WF Studies}
As overviewed in Section~\ref{sub:wf}, state-of-the-art DNN-based WF attacks achieve high accuracies even with 25 labeled samples per website. 
However, several recent works~\cite{juarez2014critical, cherubin2022online, critique, wang2016realistically, wang2020high} have criticized the relevance of such attacks in practice, as existing WF attacks lack realistic assumptions in their threat model, making them impractical in real-world.  The following are the main criticisms:

\paragraphb{Resilience to concept drift:} Concept Drift is one of the main issues making a WF classifier outdated as the content of many websites is changing everyday. Juarez et al.~\cite{juarez2014critical} show that concept drift can cause a significant drop in the accuracy of WF classifiers. 
To overcome this problem, the attacker needs to re-train the model regularly by fetching updated data~\cite{wang2016realistically}. Appendix~\ref{app:drf} elaborates on different types of concept drift that impact WF attacks.

\paragraphb{Network condition variations:} The network conditions where Tor traces are collected have different characteristics in terms of their bandwidth, latency, congestion, and so forth. A mismatch between the network conditions of traces used for training and the traces the attacker observes in deployment can affect the performance of the WF model significantly. Researchers usually collect traces in stable and consistent network conditions which is not always the case during deployment. Clients in different locations may experience low bandwidth connections or high latency affecting the underlying features of their Tor traces. 

\paragraphb{Inaccurate user imitation:} Researchers often use automated browser crawlers such as Selenium~\cite{selenium} positioned in an entry relay to collect ground truth traces for training WF classifiers. However, these synthetic traces do not reflect actual behavior of a Tor client, such as different browser configurations or visiting subpages of a website, e.g., it is impractical to assume that clients only have one tab open while clients usually visit websites concurrently~\cite{von2013dobbs, xu2018multi}. 
Researchers have investigated the effect of multi-tab browsing in WF attacks~\cite{juarez2014critical, wang2016realistically}. 
Online WF~\cite{cherubin2022online} shows that synthetic traces cannot represent genuine Tor traces. They modify the threat model of a conventional WF attack by collecting genuine Tor traces from an exit relay for training the WF classifier. 

\paragraphb{Requiring large labeled datasets:} Despite the high accuracy of DNN-based WF attacks, they often require large amounts of labeled data, e.g., DF~\cite{sirinam2018deep} uses 800 samples per website to achieve $98\%$ accuracy. Gathering labeled data in any learning-based scenario in general, and in WF attacks in particular require excessive effort making it impractical in real-world.  
Researchers used different techniques such as contrastive learning~\cite{sirinam2019triplet}, GANs~\cite{oh2021gandalf}, and residual networks~\cite{var-cnn} to mitigate the reliability of WF attacks on labeled data. 

Despite the partial success of recent WF studies to ease the challenges of such attacks in real-world scenarios, they still lack practicality when it comes to actual Tor traces. 
In this work, we aim for the root cause of the mentioned critics: lack of longitudinal perspective into network traffic. As a response, we leverage carefully designed data augmentation tailored to the Tor network to represent the diverse and volatile nature of web traffic. A Tor-tailored augmentation enables the WF model to obtain traces in unobserved settings by replicating the modifications that may happen during the deployment of the attack. 


\subsection{Adversary Model}\label{sub:adversary-model}

\begin{figure}[t]
  \centering
  \includegraphics[width=0.7\linewidth]{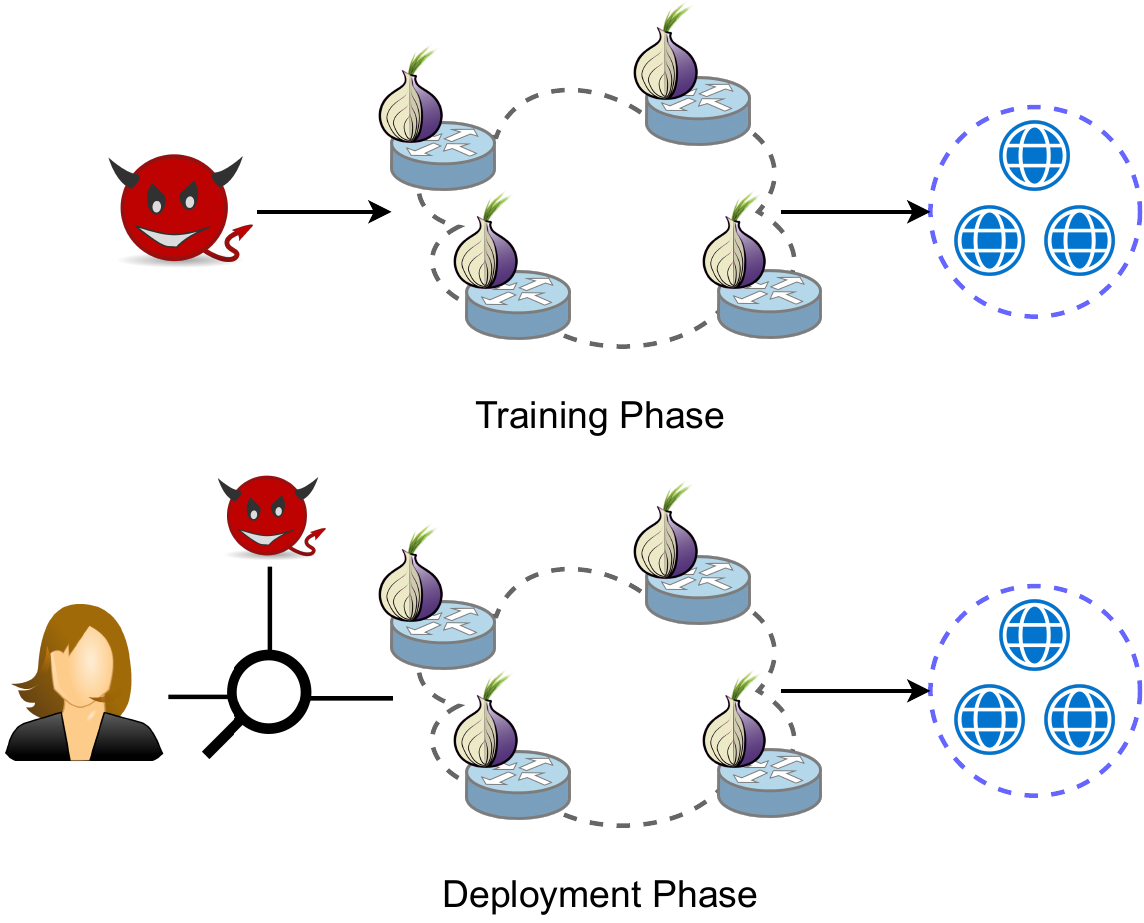}
  \caption{Setup of a machine learning-based WF attack.}
  \label{fig:wf}
\end{figure}

In this work, we consider a passive and local adversary for Tor. Network administrators, Internet Service Providers (ISP), and Autonomous Systems (AS) are examples of a local adversary having access to the link between the client and the entry relay. The adversary collects TCP packets from which it can extract Tor cells. We assume that the adversary cannot break the encryption provided by Tor.
In machine learning-based WF attacks, the adversary first trains a classification model (mostly DNNs in recent works) and then deploys the model against users' traffic. Figure~\ref{fig:wf} shows the setting of a machine learning-based WF attack. 

In this work, the attacker has the same interception point as previous WF attacks. The attacker collects traces by running an entry relay and then uses these traces to train a WF classifier. 
Note that, due to privacy reasons and to keep the anonymity of Tor users, we do not evaluate our system on  Tor traces of actual users. 
Furthermore, an evaluation on genuine Tor traces cannot be shared publicly making it challenging for future reproducibility. 

To evaluate the ability of the attacker to generalize for unobserved settings during deployment, we collect traces in different network settings for training and deployment. These traces are synthetic as we need to know the labels to train the model.
We assume that the attacker uses traces that are collected in consistent network conditions with high bandwidth and low latency. Hence, the attacker performs all the training phases on \highbw traces. 
On the other hand, during deployment, the attacker may encounter traces in various network conditions such as low bandwidth and high latency. So to consider the worst-case scenario, we evaluate the performance of our attack on \lowbw traces. 
In Section~\ref{sec:data}, we elaborate on how we divide the traces of our dataset into \lowbw and \highbw traces. 
We believe that in such a scenario, we evaluate the capability of the model of learning the underlying features of traces in unobserved settings. In Sections~\ref{sec:aug} and \ref{sec:model} we explain how we achieve this goal by leveraging data augmentation and deploying it through \self and \semi algorithms.

\section{N{\small et}A{\small ugment}: Augmenting Network Traces}\label{sec:aug}
As mentioned before, limitations in data collection are one of the main weaknesses of existing WF attacks, questioning the relevance of such attacks in practice.
We present augmentation as a potential solution to this problem as it enables the model to train on different variations of website traces which it might later face when it is deployed. These traces can have a distribution different from that of the training data, as there are unpredictable sources of noise causing variations in traffic which might not be present when training data is collected.  
An augmentation tailored to the domain of network packets and Tor cells enables the adversary to perform classification on traces in a realistic scenario.  
Augmentation also allows the attacker to extend their training dataset, therefore eliminating the need for large amounts of labeled data. 
There are numerous works focusing on augmentation techniques applied to images in computer vision. These techniques range from basic image manipulations, such as resizing, flipping, shifting, and cropping, to more complex techniques such as AutoAugment~\cite{cubuk2019autoaugment}, RandAugment~\cite{cubuk2020randaugment}, and CTAugment~\cite{Berthelot2020ReMixMatch}.
However, augmenting the network traces is more challenging due to the different nature of Tor network traces. Specifically, one cannot simply borrow data augmentation techniques designed for classical machine learning tasks. Instead, to get the most out of augmentation, we develop augmentation techniques that are tailored to the specific characteristics of network traffic. 
To come up with a data augmentation approach for Tor traces, we started by implementing a naive augmentation called \flipaug where for each Tor cell, we flip its direction with a probability of $p_{flip}$.
\flipaug is a simple augmentation that does not necessarily reflect effects of variations that can occur in Tor network conditions.

\begin{figure*}[t]
  \centering
  \includegraphics[width = 0.8\linewidth]{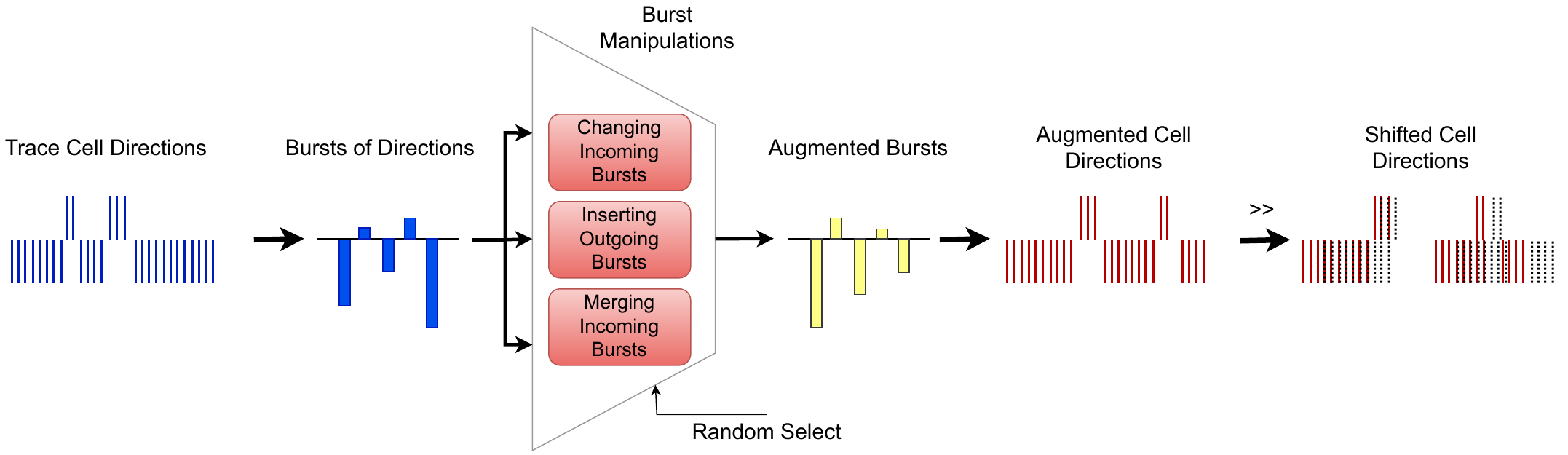}
  \caption{Overview of \aug}
  \label{fig:aug}
\end{figure*}

We then propose \emph{\aug}, a new data augmentation technique tailored to the specific characteristics of Tor network traces. We believe that using a Tor-tailored augmentation is necessary to enable the model to obtain network samples that represent unobserved network conditions and disparate settings. We demonstrate the effectiveness of \aug as opposed to \flipaug through extensive experiments in Section \ref{exp}. 
Figure~\ref{fig:aug} shows the high-level description of \aug.
\aug focuses on bursts of Tor cell directions since the traces fed to the attack models are sequences of cells represented by their direction (incoming or outgoing)
(See Section \ref{sub:format} for more details on data representation).
We define a sequence of consecutive cell directions as a burst if they all have the same direction. The number of cells in each burst is considered the size of that burst.
In a Tor trace, incoming bursts consist of cells transmitted from the website to the client and outgoing bursts consist of cells captured in the other direction.
For each trace, we first extract the incoming and outgoing bursts, then we apply one of three burst manipulations, and finally, we apply a shift transformation. Each manipulation represents one or several cumulative effects of varying network conditions on traces. 

\begin{figure}[t]
  \centering
  \includegraphics[width = 0.8\linewidth]{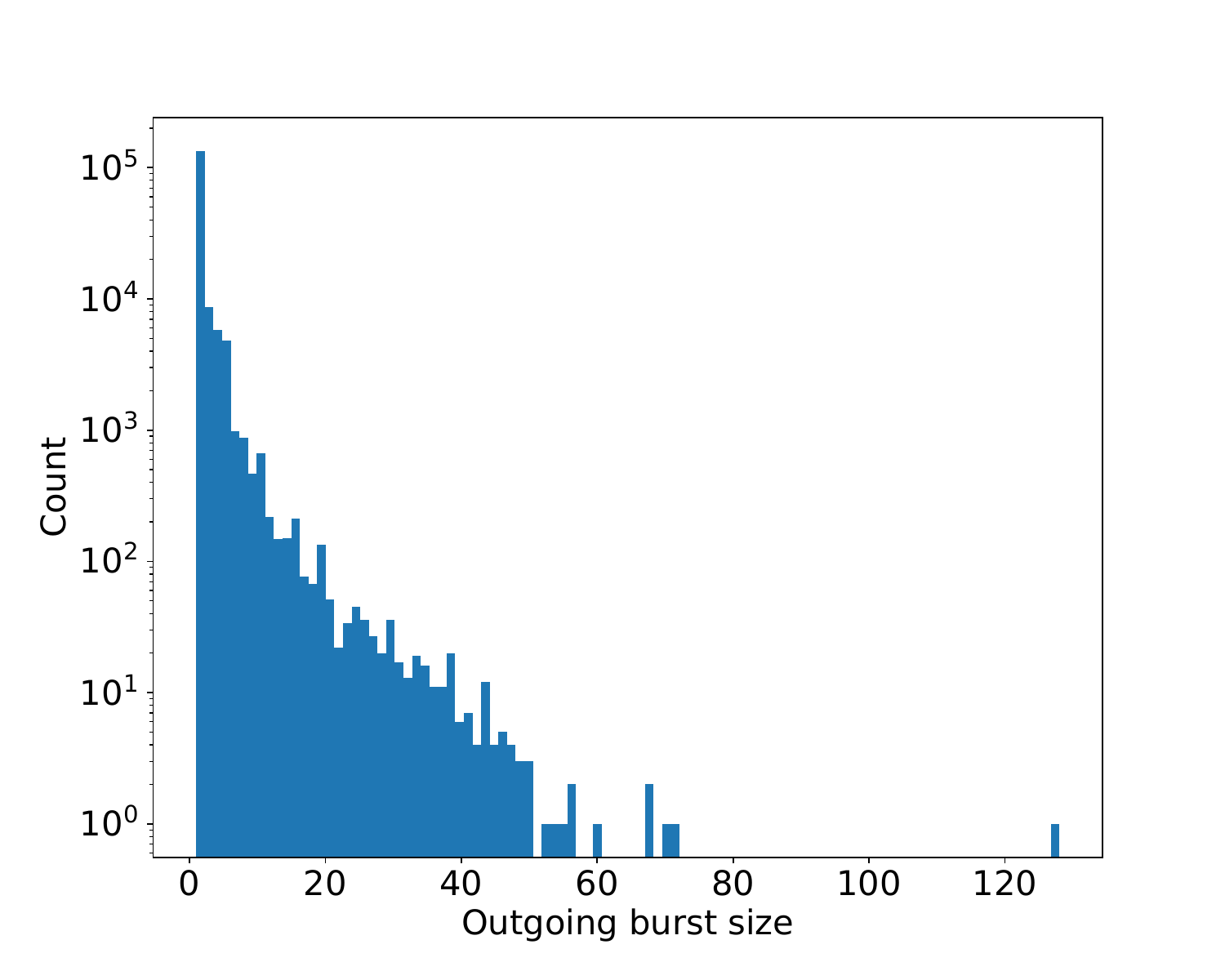}
  \caption{Empirical distribution of outgoing burst sizes from 1000 traces. To insert outgoing bursts we randomly sample from this distribution.}
  \label{fig:out-dist}
\end{figure}

Since the WF literature defines an incoming cell as -1 and an outgoing cell as +1, incoming bursts have negative sizes and outgoing bursts have positive sizes. 
Note that when applying each burst manipulation on a trace, we do not modify the first 20 cells, as the first cells are often used for the protocol initiation and handshake which means they remain the same among different traces of a particular website. 
For each trace, \aug randomly applies one of the following burst manipulations:
\begin{itemize}
    \item \textbf{Modify incoming burst sizes:} The content of most websites is changing every day and as a result, the classifier may not be able to capture the unique pattern of each website. 
    The bursts of incoming Tor cells in a trace contain downloaded contents of the websites such as text, images, and other parts of the website. 
    Figure~\ref{fig:num-incoming} shows the mean and standard deviation of the number of incoming cells in traces of 50 websites randomly chosen from AWF dataset. This figure shows how different traces of the same website can have varied numbers of incoming cells, indicating the dynamic nature of the contents of a website.
    To replicate this variation in the contents of a website, we randomly modify the size of incoming bursts and generate new network traces for the same website. 
    For traces with less than 1000 Tor cells, we only increase the size of incoming bursts and for traces with more than 4000 Tor cells, we only decrease the size of incoming bursts. For traces with a number of Tor cells between 1000 and 4000, we randomly choose to increase or decrease the size of incoming bursts. 
    The modification of incoming burst sizes happens with rates $r_{upsample}$ and $r_{downsample}$ which are the hyper-parameters of \aug. 
    
    \begin{figure}[t]
      \centering
      \includegraphics[width = 0.8\linewidth]{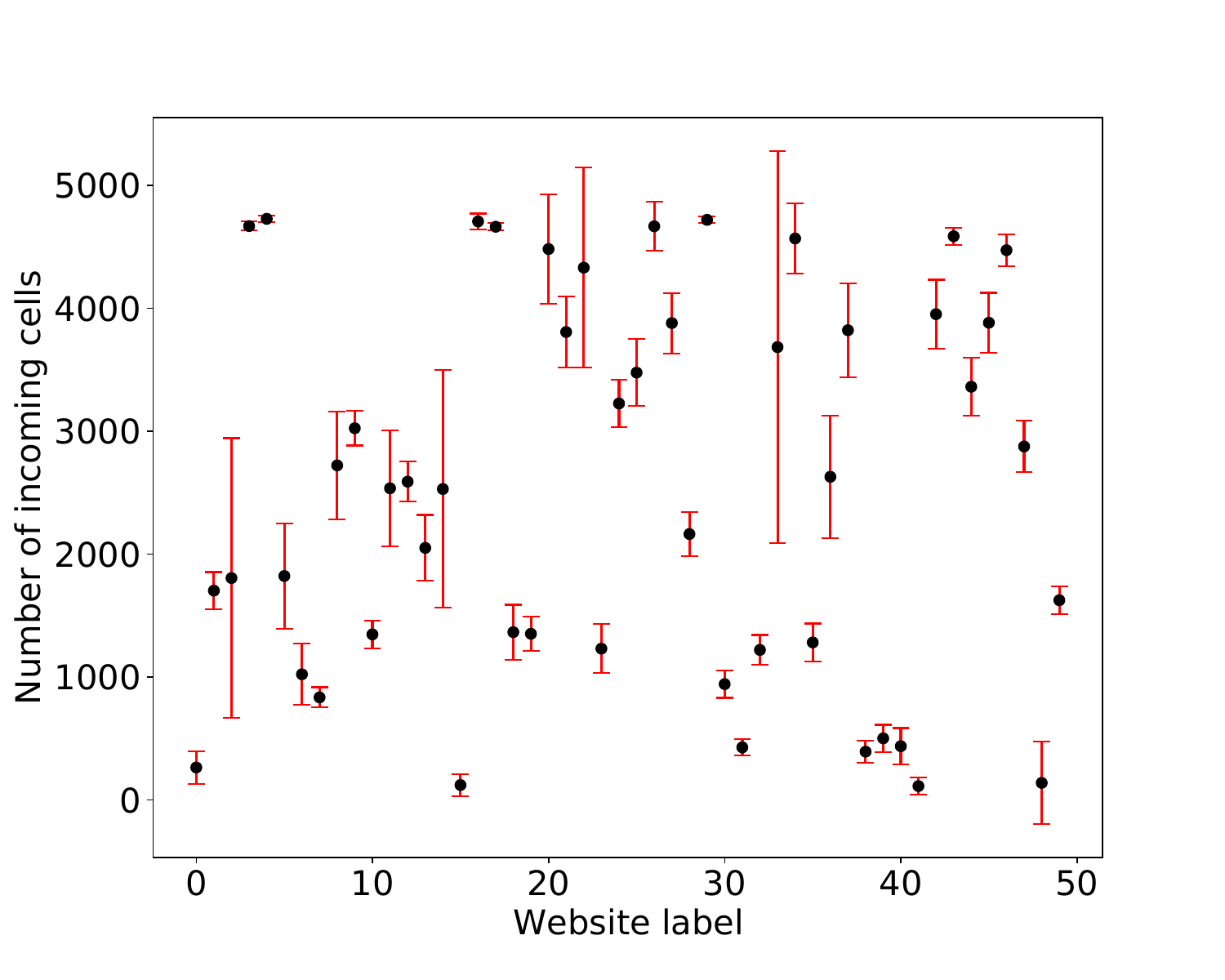}
      \caption{Mean and standard deviation of the number of incoming Tor cells in traces of 50 websites in the AWF dataset. Different samples of the same websites have a varied number of incoming Tor cells due to their dynamic content.}
      \label{fig:num-incoming}
    \end{figure}
    
    \item \textbf{Insert outgoing bursts:} Tor sends control cells periodically for flow control and other purposes. SENDME cells are the most common control cells that can affect the traffic analysis algorithm ~\cite{wang2016website}. 
    Different network conditions lead to different circuit bandwidths which affect the number of control cells that are present in each trace, e.g., when a client is connecting to a low bandwidth circuit, there could be more control cells in their network trace. 
    To represent this effect in our augmentation, we randomly split incoming bursts and insert an outgoing burst to generate an augmented network trace. 
    To choose the size of these inserted outgoing bursts, we use the empirical distributions of approximately 198k outgoing burst sizes obtained from 1000 traces of AWF dataset. Figure~\ref{fig:out-dist} shows this distribution. 
    Inserting outgoing bursts happens at a rate $r_{insert}$ which is a hyper-parameter of \aug.

    \item \textbf{Merge incoming bursts:} As mentioned previously, higher circuit bandwidths can translate into fewer control cells. Therefore, by merging the incoming bursts, our augmentation represents this variation while maintaining the amount of incoming data which is consistent among most of the traces of a single website. 
    We represent this effect by merging incoming bursts and removing some outgoing bursts randomly.
    We merge $n_{merge}$ number of incoming bursts at a rate $r_{merge}$. $n_{merge}$ and $r_{merge}$ are hyper-parameters of \aug. 
    
\end{itemize}

Once the burst manipulation is applied, the sequence of bursts is converted to a sequence of cells. Then, the last step of \aug is to shift the cells. To shift a trace by $n$ cells, we drop the last $n$ cells of the trace and insert $n$ zero-sized cells to its beginning.
When deploying the attack, the adversary observing the victim's traffic, may not know which cell is the first cell in a trace. This may cause that particular trace to be shorter than previously observed samples of that website. The intuition behind this shifting step is to represent shorter traces not just by zero-sized ending cells, but also by introducing zero-size leading cells to increase resilience to virtual concept drift (See Appendix~\ref{app:drf} for details on types of \drf in WF).  
Algorithm~\ref{alg:aug} summarizes the steps in \aug. Furthermore, Algorithms~\ref{alg:modify}, \ref{alg:merge}, and \ref{alg:insert} show the detailed implementation of the burst manipulations in \aug. 
Table~\ref{tab:aug-param} shows the optimal values of \aug hyperparameters.
We select each hyperparameter by searching through a set of candidates. We pick the hyperparameter which leads to the best accuracy.

\begin{algorithm}[!t]
\caption{\aug Algorithm}
\begin{algorithmic}
    \State $t \gets$ vector of cell directions
    \State $SHIFT \gets$ shift parameter
    \State $bursts$ = extract\_bursts($t$) 
    \State $manipulations \gets$ \{ \\
    \hspace*{4em}Modify Incoming Burst Sizes (Algorithm~\ref{alg:modify}),\newline 
    \hspace*{4em}Merge Incoming Bursts (Algorithm~\ref{alg:merge}),\newline \hspace*{4em}Insert Outgoing Bursts (Algorithm~\ref{alg:insert}) \\
    \hspace*{4em}\}
    
    \State $\mathcal{M} \gets$ Randomly pick from $manipulations$
    
    \State $bursts_{augmented} = \mathcal{M}(bursts, t)$
    
    \State $t_{augmented} \gets$ convert\_burst\_to\_cells($bursts_{augmented}$)
    
    \State $n$ = Pick random value from $\{0, \cdots, SHIFT\}$
    
    \State $output$ = $t_{augmented} >> n$
\end{algorithmic}
\label{alg:aug}
\end{algorithm}

\setlength{\floatsep}{3ex}
\setlength{\textfloatsep}{3ex}
\begin{algorithm}[!t]
\caption{Modifying Incoming Burst Sizes Algorithm}
\begin{algorithmic}
    \Function{Modify\_Size\_of\_Bursts}{$bursts$, $t$}
        \State $r_{upsample} \gets$ The rate of increasing burst size
        \State $r_{downsample} \gets$ The rate of reducing burst size
        \State $burst\_size\_threshold$ $\gets$ Minimum number of non-zero\newline
        \hspace*{6em}cells in a burst for the manipulation to be applied 
        \If {len($t \neq 0) <= 1000$}
            \State delta = $r_{upsample}$
        \ElsIf {len($t \neq 0) > 4000$}
            \State delta = -$r_{downsample}$
        \Else
            \LineComment{Randomly decide to increase or decrease size}
            \State delta = Pick random value from \{$r_{upsample}$, -$r_{downsample}$\}
        \EndIf
        \For {$burst\_size$ in $bursts$}
            \LineComment{Skipping bursts with less than 10 cells}
            \If {$burst\_size \le -burst\_size\_threshold$}
                \State $burst\_size \mathrel{\times}= (1+\text{random}[0,1]\times delta)$

            \EndIf
        \EndFor
        return $bursts$
    \EndFunction
\end{algorithmic}
\label{alg:modify}
\end{algorithm}

\begin{algorithm}[!t]
\caption{Inserting Outgoing Bursts Algorithm}
\begin{algorithmic}
    \Function{Insert\_Outgoing\_Bursts}{$bursts$}
        \State $r_{insert}$ $\gets$ The rate of inserting outgoing bursts
        \State $\mathcal{BS}$ $\gets$ Empirical distribution of burst sizes
        \For {$burst\_size$ in $bursts$}
            \If{$burst\_size < 0$}\Comment{Ignoring outgoing bursts}
                \State $random\_prob$ = $\text{random}[0,1]$
                \If {$random\_prob < r_{insert}$}
                    \State $size$ = Sample($\mathcal{BS}$)
                    \State $position$ = Pick random value  \newline
                    \hspace*{13em} from $\{3, \cdots, burst\_size-3\}$
                    \State $bursts$.insert(size=$size$, position=$position$)
                \EndIf
            \EndIf
        \EndFor
        return $bursts$
    \EndFunction
\end{algorithmic}
\label{alg:insert}
\end{algorithm}

\begin{algorithm}[!t]
\caption{Merging Incoming Bursts Algorithm}
\begin{algorithmic}
    \Function{Merge\_Incoming\_Bursts}{$bursts$}
        \State $n_{merge}$ $\gets$ Number of bursts to merge in each step
        \State $r_{merge}$ $\gets$ The rate of merging the bursts
        
        \For{$burst\_size$ in $bursts$}
            \If{$burst\_size < 0$}\Comment{Ignoring outgoing bursts}
                \State $random\_prob$ = $\text{random}[0,1]$
                \If {$random\_prob < r_{merge}$}
                    \State $num\_merges$ = Pick random value\newline
                    \hspace*{13em} from $\{2,\cdots, n_{merge}\}$
                    \State $new\_burst\_size$ $\gets$ Merging $num\_merge$ \newline 
                    \hspace*{14em} consecutive bursts
                \EndIf
            \EndIf
        \EndFor
    \EndFunction
\end{algorithmic}
\label{alg:merge}
\end{algorithm}

\begin{table}[!t]
    \centering
    \caption{Hyperparameters of \aug}
    \begin{tabular}{||c|c|c||}
        \hline
        \textbf{Parameter} & \textbf{Search Space} & \textbf{Choice} \\
        \hline
        $SHIFT$ & $\{ 5,10,20, 50\}$ & $10$\\
        \hline
        $r_{upsample}$ & $0.1 \sim 1$ & $1$\\ 
        \hline
        $r_{insert}$ & $\{0.1, 0.3, 0.5, 0.7 \}$ & $0.3$\\
        \hline
        $r_{downsample}$ & $0.1 \sim 1$ & $0.5$\\ 
        \hline
        $burst\_size\_threshold$ & $\{10, 20\}$ & $10$\\
        \hline
        $n_{merge}$ & $\{3, 4, 5, 6 \} $ & $5$\\
        \hline
        $r_{merge}$ & $\{0.05, 0.1, 0.2, 0.3 \}$ & $0.1$\\
        \hline
    \end{tabular}
    \label{tab:aug-param}
\end{table}



\section{Website Fingerprinting Using Augmented Traces}\label{sec:model}


As mentioned in the previous section, augmentation can boost the performance of WF attacks under realistic scenarios by enabling the model to obtain traces in unobserved settings and conditions.
Augmented Tor traces can be used in different methods to train a WF classifier. 
In this section, we explain how we instantiate \aug through \self and \semi techniques, and then we describe our proposed deployments of augmentation using \self and \semi: \ourWF and \netfm.
We use \semi and \self techniques to remove the requirement of gathering large labeled traces by the adversary. 

\subsection{\ourWF}

\begin{figure}[t]
  \centering
  \includegraphics[width = 0.9\linewidth]{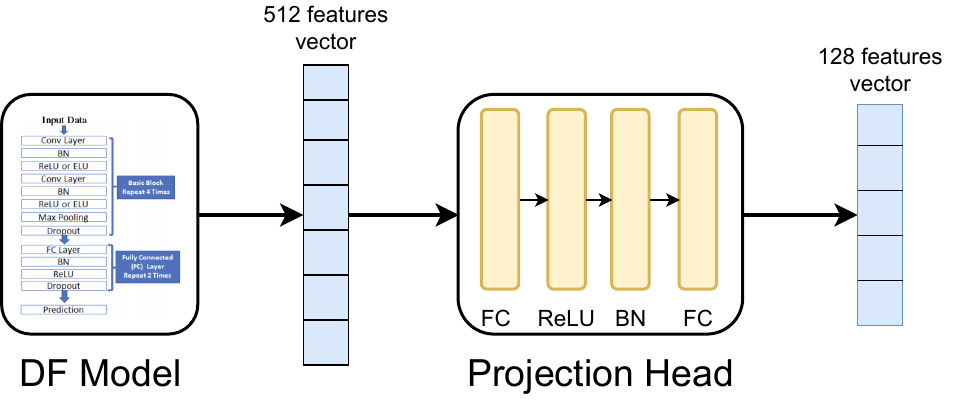}
  \caption{\ourWF Pre-train Structure}
  \label{fig:netclr}
\end{figure}

We propose \ourWF, a WF attack technique that uses contrastive learning and network trace augmentation to learn accurate representations of network traces. \ourWF is based on \self and does not require any labeled data for the pre-training phase. \ourWF adopts the methodology of SimCLR~\cite{chen2020simple} and adjusts its components to the domain of network traces. 
\ourWF consists of three phases to perform the WF attack. 

\begin{figure*}[t]
  \centering
  \includegraphics[width = 0.8\linewidth]{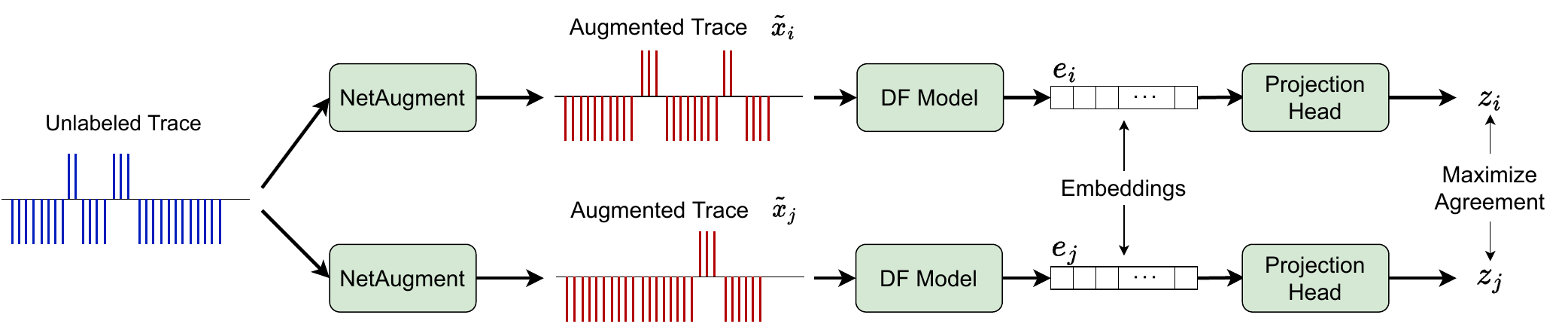}
  \caption{\ourWF pre-training steps}
  \label{fig:simclr}
\end{figure*}

\paragraphb{Pre-Training Phase: } In the pre-training phase, we train a \self model that learns to generate lower-dimension representations for website traces. This is similar to what TF~\cite{sirinam2019triplet} does, however, as opposed to TF which needs at least 25 labeled samples for each of its 775 websites, \ourWF does not need any labeled samples to perform the pre-training.
Instead, \aug helps the model to see different samples of a network trace and generate representations that are sufficiently close to each other for the same website and far from samples of other websites. 
As suggested by TF~\cite{sirinam2019triplet} and Online WF~\cite{cherubin2022online}, the base network of \ourWF is the DF neural network proposed by Sirinam et al.~\cite{sirinam2018deep}. We also add a projection head to the top of DF as it improves the performance of the pre-training~\cite{chen2020simple}. The projection head contains two fully connected layers with a ReLU activation function and a Batch Normalization~\cite{ioffe2015batch} layer.
Figure~\ref{fig:netclr} shows the structure of \ourWF pre-training model. 
The pre-training of \ourWF learns representations of network traces and converts the traces with 5000 features to representations with a length of 512. The projection head converts the representations to an output of size 128. Figure~\ref{fig:simclr} shows the \ourWF pre-training steps.  

\paragraphb{Fine-Tuning Phase: } In the fine-tuning phase of \ourWF, the adversary uses different numbers of labeled traces, denoted as $N$, to fine-tune the DF model that is pre-trained on augmented traces. In this phase, we replace the projection head with a simple fully connected layer with probabilities of the input trace belonging to each class. 
We train the whole base network plus the fully connected layer. This is similar to the semi-supervised evaluation method proposed in~\cite{chen2020simple} where the pre-trained SimCLR is fine-tuned using a small portion of the dataset as the labeled dataset. 
We consider both the pre-training and fine-tuning phases the \textit{training phase}.

\paragraphb{Deployment Phase: } Similar to all WF attacks, in the attack phase, the adversary performs the actual WF attack and uses the fine-tuned model to identify the traces visited by clients.



\subsection{\netfm}
We also instantiate \aug through \semi techniques. 
We adopt the implementation of FixMatch~\cite{sohn2020fixmatch} and integrate \aug into it. 
We then present \netfm, a WF attack based \semi and \aug that uses pseudo-labels to generate labels for the unlabeled portion of the dataset. We then train the WF classifier using the augmented traces with generated pseudo-labels. 
The backbone of \netfm algorithm is also the DF neural network~\cite{sirinam2018deep} with the same parameters. For weak augmentations, we use \flipaug with $p_{flip} = 0.1$.

\noindent Note that for both \netfm and \ourWF we use the same parameters for DF as the ones used in the original paper~\cite{sirinam2018deep}.

\section{Data Collection and Setup}\label{sec:data}



\subsection{Network Condition Metric}\label{sub:network-condition}
For evaluating a model in the realistic scenario described in Section \ref{sub:adversary-model}, we need a metric to differentiate \highbw and \lowbw traces. We define the \textit{network condition metric (NCM)} of a trace as the ratio of the total size of downstream Tor cells to the loading time of that trace. We approximate the loading time of each trace by the difference between the timestamps of it first and last cell. We believe this metric can reflect the cumulative effects of changes in bandwidth, latency, loss, and congestion in Tor relays, clients, and servers. For traces in our datasets, we found 40 kBps to be the appropriate threshold for the NCM to partition \highbw and \lowbw traces as the performance of existing WF techniques starts to drop for traces with an NCM value below this threshold.

\subsection{Model Input Representation}\label{sub:format}
We use the same representation shared by recent work in WF~\cite{rimmer2017automated, sirinam2019triplet, wang2013improved, sirinam2018deep} for model inputs. Each sample of website traffic trace used for training and testing a model is converted into a sequence of Tor cells represented by +1 and -1 for outgoing and incoming cells, respectively. Since model inputs have a fixed length of 5000, longer sequences are truncated and shorter traces are padded with zeros.
For each of the datasets described in Section \ref{sub:datasets}, these sequences are processed to filter out some traces before they are used as model inputs. Empty traces are discarded. In the closed world setting, similar to ~\cite{wang2013improved}, for each trace of a website, if its size is less than 20\% of the median trace size of the website, it is discarded.
In the open-world setting, all traces with less than 20 cells are discarded.

\subsection{Dataset Labels and Composition}\label{sub:datasets}

\paragraphb{AWF dataset.} This large dataset of non-onion websites dataset was collected by Rimmer et al.~\cite{rimmer2017automated} in 2017 using Tor browser version 6.5. By contacting the authors, we obtained the full parsed traces of the dataset which included metadata such as packet timestamps. The parsed traces we obtained include up to 3000 traces for the homepage of 1200 monitored websites as well as  traces generated by one-time visits to 565947 unmonitored websites. After processing these traces according to the method described in Section \ref{sub:format}, we categorize them into several different sets for different scenarios.
For the traditional WF scenario in the closed world setting where the NCM is not taken into account, we assemble the following: 
\begin{compactitem}
    \item \textbf{AWF1:} The set of traces for 100 randomly picked monitored websites.
    \item \textbf{AWF2:} The set of traces for another 100 randomly picked monitored websites. Note that the set of websites in AWF1 and AWF2 are distinct.
\end{compactitem}


For the scenario in the closed-world setting where the NCM is taken into account for training, we assemble the following:
\begin{compactitem}
    \item \textbf{AWF-{attack}:} The set of traces for the same 69 monitored websites from the AWF100 dataset in \cite{sirinam2019triplet} which had enough superior and inferior traces. This dataset is further split into \highbw and \lowbw traces denoted by \textbf{AWF-A\textsubscript{sup}} and \textbf{AWF-A\textsubscript{inf}}, respectively.

    \item \textbf{AWF-{pre-training}:} The set of traces for 100 other monitored websites, where the websites are randomly picked. 
    Note that the set of websites in AWF-pre-training and AWF-attack are distinct.
    This dataset is further split into \highbw and \lowbw traces denoted by \textbf{AWF-PT\textsubscript{sup}} and \textbf{AWF-PT\textsubscript{inf}}, respectively. Each website in AWF-pre-training has 500 traces in AWF-PT\textsubscript{sup} and 500 traces in AWF-PT\textsubscript{inf}.
\end{compactitem}
For the open-world setting, we assemble the following:
\begin{compactitem}
    \item \textbf{AWF-OW10k:} The set of traces for 10000\ \highbw \ and 10000\ \lowbw \ unmonitored websites.
    \item \textbf{AWF-OW50k}, \textbf{AWF-OW100k}, and \textbf{AWF-OW200k} are defined similarly to AWF-OW10k, with 50k, 100k, and 200k traces of both superior and inferior types, respectively.
\end{compactitem}

\paragraphb{Drift dataset. } We collected this dataset to study the impact of concept drift on our attacks. For the closed world setting, we collected up to 550 traces for visits to each one of 225 non-onion websites. Similar to \cite{rimmer2017automated}, this list of websites was compiled so as to avoid duplicate entries that only differ in the top-level domain as a means of website localization. The set of these monitored websites is distinct from those in AWF-pre-training. While limiting the guard relays to 18 specific relays located either in North America or Europe, we collected over 100 traces for 112 non-onion websites. The purpose of this set of traces is to investigate the effect of the location of guard relays on WF performance.
For the open-world setting, we collected a single instance for each website. We picked 10000 websites from unmonitored websites in the AWF dataset. As a result, the set of these websites is also distinct from those in AWF-pre-training. 
Note that this dataset was collected more than 5 years after the AWF dataset.
After processing these traces according to the method described in Section \ref{sub:format}, we label the subsets as:
\begin{compactitem}
    \item \textbf{\driftcw:} The set of 90 monitored websites, where each site has at least 100 \highbw and 20 \lowbw traces. This dataset is further split
    into superior and inferior traces denoted by \textbf{\driftcw\textsubscript{sup}} and
    \textbf{\driftcw\textsubscript{inf}}, respectively.
    \item \textbf{\driftguard:} The set of 90 monitored websites is further split into a set of traces collected through 11 guard relays in Europe and a set of traces collected through 7 guard relays in North America. 
    \item \textbf{\driftow:} The set of 5000 unmonitored websites.
\end{compactitem}


\subsection{Creating the Drift Dataset}\label{sub:driftdataset}


\subsubsection{Data Collection}\label{collection}
We collect traces for the Drift dataset on multiple Ubuntu 20.04 virtual machines set up through KVM over the course of three months in multiple batches.
We use the Python library tbseleinum version 0.6.3~\cite{tbseleinum} to automate the Tor browser bundle (TBB) version 11.0.10 and use Stem~\cite{stem} to interact with the controller interface of TBB's \texttt{tor} process through Python. As recommended by ~\cite{wang2013improved}, we set \texttt{UseEntryGuards} to 0 in \texttt{torrc} to disable the set of limited entry guards and disable browser caching. This makes the collected data more realistic. Using asynchronous Tor controllers, we listen for \texttt{STREAM}, \texttt{STREAM\_BW}, and \texttt{CIRC} events. While the main thread of the script is browsing different websites to collect traffic, a separate thread processes the event queue for these three event types. The \texttt{STREAM\_BW} events include the timestamp of when a specific stream was used to send and receive bytes. We listen to \texttt{CIRC} events so that every time a new circuit is created we have its timestamp, id, and path which includes relay IPs and fingerprints. \texttt{STREAM} events show which circuit each stream is attached to, as well as other information such as which website is the target of that stream. The information collected from these events was stored for later use in processing packet capture files.

In each round of collecting traffic, we open a new tab, close the previous one, and then start capturing packets using \texttt{tcpdump}. We wait 5 seconds to ensure the capture has started, then navigate to a website. Once the load event has been fired, we wait for another 15 seconds. We then stop \texttt{tcpdump} and log the consensus bandwidth file if new measurements are available. Once every website is visited, or if there has been an error, we wait until \texttt{tor} would accept a \texttt{NEWNYM} signal and then manually renew Tor circuits by sending this signal. We also restart the Tor browser to make sure that the browser cache is cleared. We also keep a log of any errors so we can discard the corresponding packet capture files when processing the data. If the script is collecting traffic for monitored websites, it is then ready to start another round after the restart at the end of the previous round.

\subsubsection{Processing}\label{processing}
Since Tor cells are embedded in TLS records, we parse the packet capture files by extracting the TLS records from TCP packets using \texttt{tshark}~\cite{tshark}. The length of a TLS record can be used to approximate the number of embedded Tor cells, as the size of a Tor cell is either 512 or 514 ~\cite{wang2013improved}. The stored \texttt{tor} event information described in Section \ref{collection}, lets us specify the IP address of the entry relay for each URL. This IP address is then used to find the relevant TLS records, discarding others. Then, the NCM is calculated for all traces before they are converted to the format described in Section \ref{sub:format}.

\subsection{Ethical Consideration}\label{sub:ethicaldata}
During the three-month period, we visited web pages to collect data for the Drift dataset, we used less than 10 clients while the Tor network had over 2 million daily users. As such, our clients should only have had a limited impact on the Tor network. Since we were collecting synthetic traffic on the same machine as the Tor clients, no information related to genuine traces was collected.

\section{Experiment Results}\label{exp}

\subsection{Experiment Setup}
We perform our experiments using PyTorch 1.12.1 and Python 3.7.
We use a single 2080 Ti GPU for all of our experiments. 
We fetch the code of existing models, DF, TF, and \gandalf, and re-run their experiments to enable a benchmark for a fair comparison. 
For the \netfm evaluation, we set $\lambda_{u} = 1$ and for the weak augmentation, we use $p_{flip}=0.1$. 
Table~\ref{tab:param} in Appendix~\ref{app:param} shows the hyperparameters for both \ourWF and \netfm. 

The following are the existing state-of-the-art techniques that we compare \ourWF to. We give brief explanations for each technique in Section~\ref{back}.
We adopt the original implementations provided by the researchers and convert them to PyTorch implementations. We made a few modifications when necessary for the data loading pipeline and hyperparameter tuning. 
\begin{compactitem}
    \item \textbf{Deep Fingerprinting (DF)~\cite{sirinam2018deep}}: DF uses convolutional neural networks to design a WF classifier that achieves $98\%$ accuracy in a traditional closed-world scenario.
    
    \item \textbf{Triplet Fingerprinting (TF)~\cite{sirinam2019triplet}}: TF adopts triplet networks to design a feature extractor that generates fixed-size embeddings for network traces. It then applied a simple KNN model to generated embeddings when there is a low number of labeled samples. Then, for a small number of labeled samples, the feature extractor generates embeddings which are then used to train a KNN model.
    
    \item \textbf{GAN for Data-Limited Fingerprinting (\gandalf)~\cite{oh2021gandalf}}: \gandalf uses generative adversarial networks to generate "fake" network traces to achieve high accuracies in limited labeled data scenario. 
\end{compactitem}

Note that since the code and dataset for Online WF~\cite{cherubin2022online} are not publicly available due to privacy reasons, we are not able to compare \ourWF with Online WF. 

\subsection{Closed-World Scenario}\label{sec:cw}
We evaluate \ourWF in a closed-world scenario where we assume that the clients only browse the set of monitored websites the adversary is interested in. 

\paragraphb{Metric:} To evaluate the performance in a closed-world scenario, we use \textit{Accuracy} which is simply the ratio of correct predictions to the total number of traces.

\subsubsection{Traditional WF Scenario}
First, we evaluate \ourWF and \aug in a traditional WF scenario where the attacker uses traces collected in the same settings for both training and evaluation. 
To perform the experiments, we use AWF2 dataset to pre-train \ourWF and we use AWF1 for the fine-tuning and evaluation data. To have a fair comparison, we use AWF2 for the feature-extraction step of TF and the unlabeled dataset of \gandalf. 
There is no pre-training phase in \netfm, hence for \netfm we use the AWF1 dataset for both the labeled and unlabeled samples.

We train each classifier using $N=\{5,10,20,90\}$ training samples per website, randomly sampled from the AWF1 dataset.
We set $\mu$ in \netfm to 19 for this experiment, e.g., in the event of 5 labeled samples, we have $19\times5 = 95$ unlabeled samples per each website. Note that this number of unlabeled traces is significantly smaller than 2500 traces used to train \gandalf. 
The test traces are chosen from AWF1 dataset and they are mutually exclusive from the samples used for training (AWF2). The test set contains 417 samples per website of the AWF1 dataset. This is consistent with the numbers used in \gandalf.  
Table~\ref{tab:limited-data} compares \ourWF with other WF techniques for different values of $N$.
Since the training examples are chosen randomly, we run each training 5 times and report the average and standard deviation of the accuracies. 
As illustrated in this table, for all values of $N$, \ourWF has a significantly higher performance than other techniques, e.g., when $N = 5$, TF can only achieve $78\%$ average accuracy while \ourWF have $89.7\%$ average accuracy. These numbers suggest that our tailored augmentation effectively helps the model learn accurate representations of website traces even with only 5 labeled examples. 
The results show that \textbf{\ourWF has higher accuracy than existing techniques for different numbers of labeled samples in the traditional WF scenario. }

\paragraphb{Effect of Augmentation}: We also evaluate \ourWF when instead of a \aug, we use \flipaug with $p_{flip} = 0.1$. 
Table~\ref{tab:limited-data} contains the results of \flipaug compared to \aug. As expected, \aug performs better than \flipaug for all the values of $N$ indicating that \textbf{a tailored augmentation is necessary to accurately replicate the unobserved settings of Tor traces.}
Note that even with \flipaug, \ourWF performs better than other systems and this is due to the promising performance of the SimCLR algorithm which is the base of the \ourWF. Specifically, even randomly flipping the directions of cells provides a weak representation of unobserved settings of Tor Traces. 

\begin{table*}
    \centering
    \caption{Comparing the performance of \ourWF with DF, TF, \gandalf, and \netfm with 5-90 labeled traces. We also compare \ourWF with a scenario where \aug is replaced with \flipaug. For all the scenarios, \ourWF outperforms other techniques. All numbers are $\%$. We do not show standard deviations less than $1\%$.}
    \begin{tabular}{@{}l|llllll@{}}
        \toprule
        N & DF~\cite{sirinam2018deep} & TF~\cite{sirinam2019triplet} & GANDaLF~\cite{oh2021gandalf} & \netfm & \ourWF \footnotesize{(\flipaug)} &\ourWF \\
        \cmidrule(r){1-1}\cmidrule(lr){2-7}
        5 & $ 60.9 \pm 2 $ & $78 \pm 1$ & $70 \pm 2$ & $77.8 \pm 1$ & $80.7 \pm 1.2$ & $\bm{89.7}$\\
        10 & $ 78.1 \pm 1.1$ & $81.6$ &  $81.1 \pm 1$ & $87.1$ & $90.5$ & $\bm{94.5}$\\
        20 & $86.1$ & $83.1$ &  $87 \pm 1$ & $93.3$ & $94.4$ &  $\bm{96.6}$\\
        90 & $96$ & $84.2$ & $95 \pm 1$ & $97.6$ & $97.7$&  $\bm{98.5}$ \\
        \bottomrule
    \end{tabular}
    \label{tab:limited-data}
\end{table*}

\subsubsection{Realistic WF Scenarios.}

In this part, we evaluate \ourWF as well as the existing WF techniques in scenarios where the Tor traces for training and evaluation are collected in different settings. 
Specifically, we perform experiments in 4 different configurations where the training phase of the WF attack happens in a setting different than the setting in the deployment phase.
Note that since \ourWF has a better performance than \netfm, for the rest of the experiments we focus on evaluating \ourWF. 

\paragraphb{Similar distributions but mutually exclusive datasets: }
The most resource intensive step in these attacks is when the adversary needs to collect a large dataset to train its model. For \ourWF and TF, it would be the dataset used for pre-training and fine-tuning, and for DF, it would be the labeled training dataset. In this scenario, we assume that the adversary only collects \highbw traces in this step. This means \lowbw traces will only be present in the deployment phase. 
We use the AWF-PT\textsubscript{sup} dataset for the pre-training phase of both \ourWF and TF. 
In the fine-tuning phase, we randomly sample $N$ labeled traces from the training subset of AWF-A\textsubscript{sup}. Lastly, in the deployment phase, we use the remaining traces in the AWF-{attack} dataset 
to generate validation, and test sets with an equal number of samples from AWF-A\textsubscript{sup} and AWF-A\textsubscript{inf} such that there are 50 \highbw and 50 \lowbw samples per website in each of the validation and test sets. The validation set is used to tune the model's hyperparameters.
To compare \ourWF with DF, we consider 3 configurations:

\begin{itemize}
    \item DF is trained only on $N$ labeled traces per website. 
    \item \dfaug is trained on augmented traces using \aug. 
    For each value of $N$, we augment the labeled traces such that there are 500 traces per website which is similar to the size of AWF-{pre-training} dataset that is used in \ourWF. This is due to enabling a benchmark for a fair comparison.   
    In particular, we want to show how much benefit we get just from \aug without modifying the training procedure (pre-training and fine-tuning of \ourWF) and network architecture. 
    \item \dfsame is trained on similar amounts of labeled data as both pre-training and fine-tuning data of \ourWF. To do so, we combine AWF-{pre-training} and AWF-{attack} datasets and train DF on them in a supervised manner. For a fair comparison with other models, we only test DF on AWF-{attack} in this configuration.
    
\end{itemize}

Table~\ref{tab:mutu} shows the comparison between different techniques when the classifier is trained only on the \highbw traces. 
As illustrated in the table, \ourWF outperforms other models significantly when the attack is performed on \lowbw traces, e.g., with 10 labeled samples for each website, TF can only achieve $64.4\%$ accuracy while \ourWF have $86.1\%$ accuracy on \lowbw traces.
Comparing DF and \dfaug shows that augmenting the traces using \aug improves the performance of DF independent of the sophisticated training procedure used in \ourWF. This indicates that \aug is beneficial on its own in that it can extend the dataset and make the DF model perform better on unobserved traces. However, using \aug combined with the training procedure of \ourWF, improves the performance of the WF attack even further, particularly on \lowbw traces. 

Furthermore, the results show that \ourWF reaches higher accuracies compared to \dfsame on both \lowbw and \highbw traces for all values of $N$. 
Note that another advantage of \ourWF compared to all configurations of DF is that the adversary does not require any labeled traces to perform pre-training as opposed to DF where the adversary needs a huge labeled dataset. 
\textbf{In summary, the results show that compared to other techniques, \ourWF is resilient to unobserved settings of Tor traces that may be present during attack deployment, even with a limited number of labeled training samples.}

Furthermore, in Appendix~\ref{app:time}, we compare the time required to train \dfsame and \ourWF. The results show that the \emph{adversary can train \ourWF faster than \dfsame by two orders of magnitude.}

\begin{table*}
    \centering
    \caption{Comparing the accuracy of \ourWF with DF and TF over \lowbw and \highbw traces in a realistic WF scenario when the distribution on training data and test data are similar. \ourWF outperforms both DF and TF on \lowbw and \highbw for different numbers of labeled samples. All numbers are $\%$. We do not show standard deviations less than $1\%$.}
    \resizebox{\textwidth}{!}{
    \begin{tabular}{@{}l|ll|ll|ll|ll|ll@{}}
        \toprule
         & DF~\cite{sirinam2018deep} & & DF \footnotesize{(Trained on augmented traces)} & & DF \footnotesize{(Trained on AWF-{pre-training} and AWF-{attack})} & & TF~\cite{sirinam2019triplet} & & \ourWF \\
        \cmidrule{2-11}
        N & \Lowbw & \Highbw & \Lowbw & \Highbw & \Lowbw & \Highbw & \Lowbw & \Highbw & \Lowbw & \Highbw  \\
        \cmidrule(r){1-1}\cmidrule(lr){2-3}\cmidrule(lr){4-5}\cmidrule(lr){6-7}\cmidrule(lr){8-9}\cmidrule(lr){10-11}
        5 &  $47.7 \pm 4.9$ & $55.3 \pm 6.2$ & $65.5$ & $80.2$ & $40.4 \pm 1$ & $55.2 \pm 1.9$ & $64.4$ & $77.9$ & $\bm{80.2}$ & $\bm{90.9}$ \\
        
        10 & $64.6 \pm 1.4 $ & $77.8 \pm 2$ & $72.9$ & $88.3$ & $53.5 \pm 1$ & $71.6 \pm 1.1$ & $69.1$ & $83.3$ & $\bm{86.1 \pm 1.2}$ & $\bm{94.8}$  \\
        
        20 & $73.6$ & $86.9$ & $77.3$ & $92.6$ & $63.6 \pm 1.1$ & $81.7$ & $73.9$ & $87.8$ & $\bm{87.1}$& $\bm{96.1}$ \\
        
        90 & $84.6$ & $93.8$ & $83$ & $95.9$ & $77.5$ & $92.5$ & $79.2$ & $92.5$ & $\bm{92.6}$ & $\bm{98}$ \\
        
        150 & $86.6$ & $94.4$ & {$85.1$} & {$96.9$} & {$80.2$} & {$94.5$} & {$79.7$} & {$93.0$} & {$\bm{93.7}$}  & {$\bm{98.1}$} \\
        
        {300} & {$89.6$} & {$95.0$} & {$87.1$} & {$97.6$} & {$83.2$} & {$96.1$} & {$81.4$} & {$94.1$} & {$\bm{94.9}$} & {$\bm{98.5}$} \\
        
        {500} & {$90.5$} & {$95.3$} & {$90.5$} & {$95.3$} & {$85.2$} & {$96.7$} & {$82.8$} & {$94.1$} & {$\bm{95.2}$} & {$\bm{98.6}$} \\
        \bottomrule
    \end{tabular}
    }
    \label{tab:mutu}
\end{table*}

\paragraphb{Comparing the effect of NCM on \ourWF:}
To investigate the effect of \highbw traces in the training phases of \ourWF, we only use \lowbw traces in the pre-training and fine-tuning phases of \ourWF algorithm. 
We then test the model trained on \lowbw traces on both \lowbw and \highbw traces. 
We perform the pre-training using AWF-PT\textsubscript{inf}.  
In the fine-tuning phase, we randomly sample $N=\{5,10,20\}$ labeled traces from the training subset of AWF-A\textsubscript{inf}. Lastly, in the deployment phase, we use the remaining traces in the AWF-{attack} dataset
to generate validation, and test sets with an equal number of samples from AWF-A\textsubscript{sup} and AWF-A\textsubscript{inf} such that there are 30 \highbw and 30 \lowbw samples per website in each of the validation and test sets. 
Table~\ref{tab:low-train} illustrates the comparison of \ourWF performance when it is trained on \lowbw and \highbw traces. We run each experiment 5 times and report the average and standard deviation of the accuracy. 
When the model is pre-trained and fine-tuned on \lowbw traces, the performance of \ourWF is better on \lowbw traces. 
However, the difference between the performance on \lowbw and \highbw traces is less when the model is trained on \lowbw traces. For instance, for $N = 10$ the \lowbw trained model has $90.9\%$ average accuracy on \lowbw traces and $86.2\%$ average accuracy on \highbw traces with a $\sim 5\%$ difference in the accuracy. On the other hand, when the model is trained on \highbw traces, the difference between accuracies is $\sim 9\%$.
This implies that when trained in the more challenging setting, the model achieves a better ability to infer underlying features of Tor cells in unobserved settings as opposed to the model trained only on \highbw traces. 

\begin{table}
    \centering
    \caption{Comparing \ourWF when the model is pre-trained on either \lowbw or \highbw traces. Training on \lowbw traces reduces the difference in the performance on \lowbw and \highbw traces. All numbers are $\%$.}
    \resizebox{\columnwidth}{!}{
    \begin{tabular}{@{}l|lll|lll@{}}
        \toprule
         & \multicolumn{3}{l|}{Trained on \lowbw traces} &  \multicolumn{3}{l}{Trained on \highbw traces} \\
        \cmidrule{2-7}
        N & \Lowbw & \Highbw & Difference & \Lowbw & \Highbw & Difference \\
        \cmidrule(r){1-1}\cmidrule(lr){2-4}\cmidrule(lr){5-7}
        5 & $85.4$ & $81.4$ & $4$ & $80.6$ & $90.1$ & $9.5$ \\
        10 & $90.9$ & $86.2$ & $4.7$ & $86.4$ & $95.1$ & $8.7$ \\
        20 & $94.2$ & $89.1$ & $5.1$ & $86.8$ & $96.7$ & $9.9$\\
        \bottomrule
    \end{tabular}
    }
    \label{tab:low-train}
\end{table}

\paragraphb{Effect of \drf:}
In this scenario, we evaluate the robustness of \ourWF against \drf. 
We use a dataset with a different distribution from AWF-{pre-training} to perform fine-tuning and evaluation. In other words, we replace AWF-{attack} with \drf.
As mentioned previously, we pre-trained \ourWF with 100 websites of the AWF-PT\textsubscript{sup} dataset collected in 2017. 
We then evaluate the pre-trained \ourWF against \driftcw dataset. There is a 5-year time gap between AWF and our collected dataset. 
For the fine-tuning and deployment phases, we use the \driftcw dataset, which consists of both \lowbw and \highbw traces, to generate training, validation, and test sets with an equal number of samples from \driftcw\textsubscript{sup} and \driftcw\textsubscript{inf} such that there are 20 \highbw and 20 \lowbw samples per website in each of the validation and test sets.
Table~\ref{tab:diff-dist} shows the results of \ourWF when fine-tuned with different numbers of labeled samples, $N$, from \driftcw. 
As Table~\ref{tab:diff-dist} illustrates, \ourWF outperforms the other techniques evaluated on both \lowbw and \highbw traces. 
As expected, the overall results are worse than the previous experiment due to the \drf effect. However, \ourWF has significantly better performance than other systems in this scenario. The results suggest that \textbf{using \aug makes \ourWF more resilient to \drf and helps the classifier to perform better against potential modifications that can happen as a result of \drf in unobserved settings during the attack}, e.g., for $N = 20$, \ourWF achieves $72.1\%$ accuracy on \lowbw traces while TF and DF can only reach to $51\%$ and $45.6\%$ respectively.

\begin{table}
    \centering
    \caption{Comparing the accuracy of \ourWF with DF and TF over \lowbw and \highbw traces in a realistic WF scenario in the presence of concept drift. The distribution of training data and test data is different. \ourWF outperforms both DF and TF on \lowbw and \highbw for different numbers of labeled samples. All numbers are $\%$.}
    \resizebox{\columnwidth}{!}{
    \begin{tabular}{@{}l|ll|ll|ll@{}}
        \toprule
         & DF~\cite{sirinam2018deep} & & TF~\cite{sirinam2019triplet} & & \ourWF \\
         \cmidrule{2-7}
        N & \Lowbw & \Highbw & \Lowbw & \Highbw & \Lowbw & \Highbw \\
         \cmidrule(r){1-1}\cmidrule(lr){2-3}\cmidrule(lr){4-5}\cmidrule(lr){6-7}
        5 & $25.2 \pm 2.3$ & $40.4 \pm 4.8$ & $41.1$ & $60.8 \pm 1.5$ & $\bm{56.2}$ & $\bm{84.4}$  \\
        10 & $36.6 \pm 1.5$ & $56.9 \pm 2.0$ & $47.0 \pm 1.4$ & $68.9$ & $\bm{66.6}$ & $\bm{92.7}$ \\
        20 & $45.6$ & $ 72.8$ & $51.0$ & $75.0$ & $\bm{72.1}$ & $\bm{96.0}$  \\
        90 & $61.9$ & $92.6$ & $56.2$ & $84.8$ & $\bm{79.6}$ & $\bm{98.3}$ \\
        \bottomrule
    \end{tabular}
    }
    \label{tab:diff-dist}
\end{table}

Furthermore, in Appendix~\ref{app:drift}, we analyze the actual observed concept drift between \driftcw and AWF-{attack} datasets by calculating the difference between their accuracy. The results in Table~\ref{tab:drift-analyze} show that the degradation in accuracy caused due to concept drift is less for \ourWF compared to DF and TF confirming that \ourWF is more resilient against concept drift.


\paragraphb{Effect of guard relay diversity:}
In this part, we evaluate the performance of \ourWF when the guard relays used to collect traces for fine-tuning and testing are mutually exclusive. 
To this aim, we use a subset of \driftguard traces that are collected through 11 guard relays located in Europe for training. The remaining traces that were collected through guard relays located in North America were used to generate validation and test sets such that there are 65 traces per website in each set.

Table~\ref{tab:guard} shows the performance of \ourWF as well as DF and TF in this setting. 
As expected, the performance of all WF techniques is worse when test traces are collected using different guard relays than fine-tuning traces. 
However, \ourWF still outperforms the other techniques in this setting, e.g., when we have 20 labeled traces, TF only achieves $59.4\%$ average accuracy when evaluated on traces with different guard relays while \ourWF has a $80.6\%$ average accuracy. 
These results confirm that \ourWF is more resilient than TF and DF to varying conditions in the Tor network that were not observed during training. 

To provide further evidence that \ourWF is resilient to previously unobserved variations in Tor traces during deployment, we also investigate effect of guard relay bandwidth. The relevant results are included in Appendix~\ref{app:consensus}.

\begin{table}
    \centering
    \caption{Comparing NetCLR with DF and TF when the guard relays used for collecting fine-tuning and testing traces are mutually exclusive. \ourWF outperforms other models when faced with unobserved guard relays. All numbers are $\%$. We do not show standard deviations less than $1\%$.}
    \resizebox{\columnwidth}{!}{
    \begin{tabular}{@{}l|ll|ll|ll@{}}
        \toprule
         & DF~\cite{sirinam2018deep} & & TF~\cite{sirinam2019triplet} & & \ourWF \\
         \cmidrule{2-7}
        N & Same & Different & Same & Different & Same & Different \\
         \cmidrule(r){1-1}\cmidrule(lr){2-3}\cmidrule(lr){4-5}\cmidrule(lr){6-7}
        5 & $43.5 \pm 2.7$ & $36.9 \pm 1.3$ & $57.5$ & $47.8 \pm 1.1$ & $\bm{71.5 \pm 1}$ & $\bm{61.3 \pm 2}$  \\
        10 & $55.5 \pm 1.1$ & $47.1 \pm 1$ & $63.9$ & $54.5$ & $\bm{82}$ & $\bm{73.4}$ \\
        20 & $67.6$ & $58.8$ & $69.7$ & $59.4$ & $\bm{87.3}$ & $\bm{80.6}$  \\
        90 & $83.2$ & $75.6$ & $77$ & $67.1$ & $\bm{93.1}$ & $\bm{89.2}$ \\
        \bottomrule
    \end{tabular}
    }
    \label{tab:guard}
\end{table}

\subsection{Open-World Scenario}

In the previous parts, we explored the performance of \ourWF in a closed-world scenario where the adversary is interested in a limited set of websites that Tor's clients are visiting. However, this is not a practical scenario in that websites users browse are not limited and they can visit any website among the huge number of websites on the Internet. 
In this section, we consider the open-world scenario, a more practical one where the adversary not only classifies traffic traces based on a limited set of monitored websites but must also distinguish whether the trace comes from a monitored set or an unmonitored one. 
Note that similar to the closed-world scenario, to evaluate \ourWF in a realistic setting we perform both the pre-training and fine-tuning using only superior traces.

For the open-world evaluation, we use the same pre-trained model in the closed-world scenario. For the fine-tuning part, as well as the monitored websites, we use a dataset of unmonitored websites that has an equal size to the monitored websites, e.g., with 10 labeled samples for each website, we have $10\times 69 = 690$ monitored traces.
We also evaluate the robustness of \ourWF against the \drf effect using our own collected dataset. 

\paragraphb{Metrics: }Since there are far more unmonitored websites and this makes the dataset imbalanced, we use Precision (P) and Recall (R) (used in WF literature~\cite{rimmer2017automated, sirinam2018deep, sirinam2019triplet}) to evaluate the performance of \ourWF in the open world scenario. 
In particular, we use prediction probabilities to compute Precision and Recall. If the input trace is a monitored website trace and the maximum output probability belongs to any monitored site and is greater than a threshold, we consider this a true positive sample. If we select the threshold such that the classifier has high precision we tune the model for precision and, if we choose the threshold for high recall we tune the model for recall. We also evaluate the models using $f_1$ score which is a weighted average of the precision and recall.  
Note that if the trace is determined to be monitored, the adversary can use the multi-class classification to identify the website the user has actually browsed.

\paragraphb{Similar distribution but mutually exclusive datasets: }
In this part, the training and evaluation datasets are from the same distribution, AWF dataset. Here we use AWF-OW10k as the unmonitored dataset and AWF-attack for monitored websites. 
Tables \ref{tab:ow-recall} and \ref{tab:ow-prec} compare the open-world performance of \ourWF with TF and DF classifiers with different numbers of labeled examples for the fine-tuning part ($N$) when the models are tuned for recall and precision respectively. 
Also, we only present the results against \lowbw traces as it is the more realistic scenario in a WF attack. 
As shown in these tables, when the model is tuned for precision, \ourWF outperforms both DF and TF significantly specifically when the attack is performed on inferior traces, e.g., using $N = 10$ labeled samples for training the classifier, TF has $48.3\%$ $F_1$ score while \ourWF achieves $77.9\%$. 
When the model is tuned for recall (Table~\ref{tab:ow-recall}), the results show that DF has a better recall for all the values of $N$; however, both DF and TF do not reach reasonable values of precision. On other hand, for all the values of $N$, \ourWF has a higher $F_1$ score than the other two techniques, e.g., when $N = 20$, although DF has $88.5\%$ recall compared to $82.5\%$ recall of \ourWF, the $F_1$ score for \ourWF is $84.8\%$ while DF has only $63.7\%$ $F_1$ score.
The overall results show that \textbf{\ourWF outperforms other systems in an open-world scenario while the model is evaluated on unobserved settings.}


\begin{table*}
    \centering
    \caption{Comparing precision and recall of \ourWF to TF and DF over \lowbw and \highbw traces when all models are tuned for recall. All numbers are $\%$.}
    \resizebox{0.7\textwidth}{!}{
    \begin{tabular}{@{}l|lll|lll|lll@{}}
        \toprule
        & DF~\cite{sirinam2018deep} & & & TF~\cite{sirinam2019triplet} & & & \ourWF \\ 
        \cmidrule{2-10}
        N & Precision & Recall & $F_1$ score & Precision & Recall & $F_1$ score & Precision & Recall & $F_1$ score \\
        \cmidrule(r){1-1}\cmidrule(lr){2-4}\cmidrule(lr){5-7}\cmidrule(lr){8-10}
        5 & $43.0$ & $77.1$ & $55.2$ & $48.9$ & $74.8$ & $59.1$ & $81.7$ & $64.6$ & $\textbf{72.2}$  \\
        10 & $44.5$ & $90.4$ & $59.6$ & $38.5$ & $78.3$ & $51.6$ & $85.0$ & $73.6$ & $\textbf{78.9}$ \\
        20 & $49.7$ & $88.5$ & $63.7$ & $40.4$ & $80.3$ & $53.8$ & $87.3$ & $82.5$ & $\textbf{84.8}$ \\
        90 & $70.2$ & $91.8$ & $79.6$ & $59.2$ & $82.7$ & $69.0$ & $90.9$ & $89.3$ & $\textbf{90.1}$ \\
        \bottomrule
    \end{tabular}
    }
    \label{tab:ow-recall}
\end{table*}

\begin{table*}
    \centering
    \caption{Comparing precision and recall of \ourWF to TF and DF over \lowbw and \highbw traces when all models are tuned for precision. All numbers are $\%$.}
    \resizebox{0.7\textwidth}{!}{
    \begin{tabular}{@{}l|lll|lll|lll@{}}
        \toprule
        & DF~\cite{sirinam2018deep} & & & TF~\cite{sirinam2019triplet} & & & \ourWF \\ 
        \cmidrule{2-10}
        N & Precision & Recall & $F_1$ score & Precision & Recall & $F_1$ score & Precision & Recall & $F_1$ score \\
        \cmidrule(r){1-1}\cmidrule(lr){2-4}\cmidrule(lr){5-7}\cmidrule(lr){8-10}
        5 & $75.8$ & $21.8$ & $33.9$ & $61.5$ & $44.5$ & $51.6$ & $92.6$ & $55.3$ & $\textbf{72.2}$ \\
        10 & $59.3$ & $55.6$ & $57.4$ & $42.7$ & $55.5$ & $48.3$ & $91.9$ & $67.6$ & $\textbf{77.9}$ \\
        20 & $60.1$ & $70.1$ & $64.7$ & $43.7$ & $63.4$ & $51.7$ & $92.7$ & $78.1$ & $\textbf{84.8}$ \\
        90 & $76.6$ & $86.8$ & $81.4$ & $67.5$ & $71.1$ & $69.3$ & $94.5$ & $86.7$ & $\textbf{90.4}$ \\
        \bottomrule
    \end{tabular}
    }
    \label{tab:ow-prec}
\end{table*}

        
        
\begin{figure}[t]
  \centering
  \includegraphics[width = 0.9\linewidth]{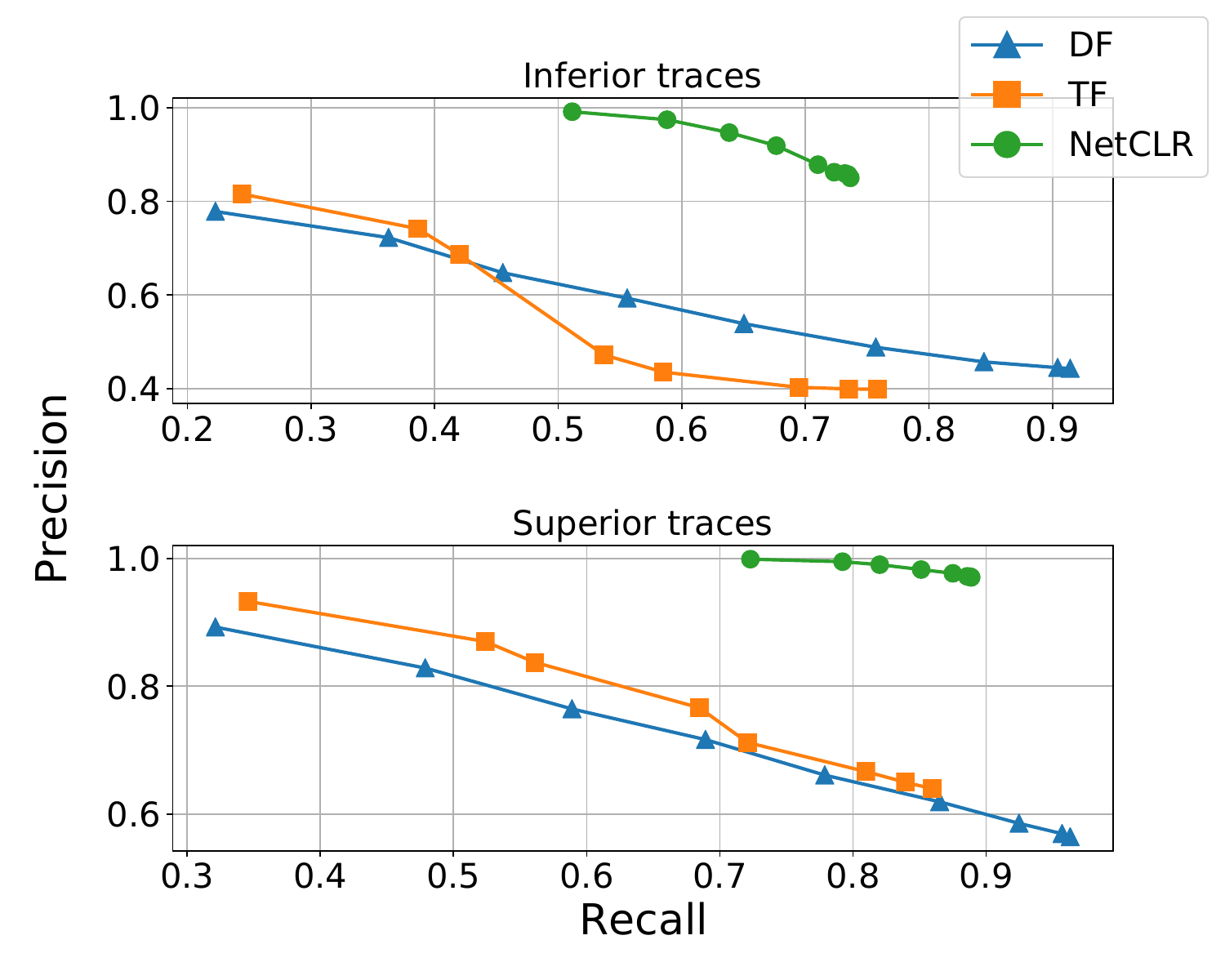}
  \caption{Comparing precision and recall of \ourWF with DF and TF over \lowbw and \highbw traces. \ourWF has better $F_1$ score comparing to other techniques.}
  \label{fig:ow-comp}
\end{figure}

We also compared \ourWF with both TF and DF with different thresholds using 10K unmonitored samples when the model is fine-tuned using 10 labeled samples per website. Figure~\ref{fig:ow-comp} shows the precision-recall curves of all three models. As expected, the performance on \highbw traces is better than \lowbw  since the model has not seen \lowbw samples during training. 
\ourWF has significantly higher precision compared to other attacks indicating that our attack rarely identifies an unmonitored site as a monitored one. 
For smaller thresholds, DF has better recall compared to other attacks but with very low precision. Overall, for all  thresholds, \ourWF has a higher $F1$ score compared to other systems.

\begin{figure}[t]
  \centering
  \includegraphics[width = 0.9\linewidth]{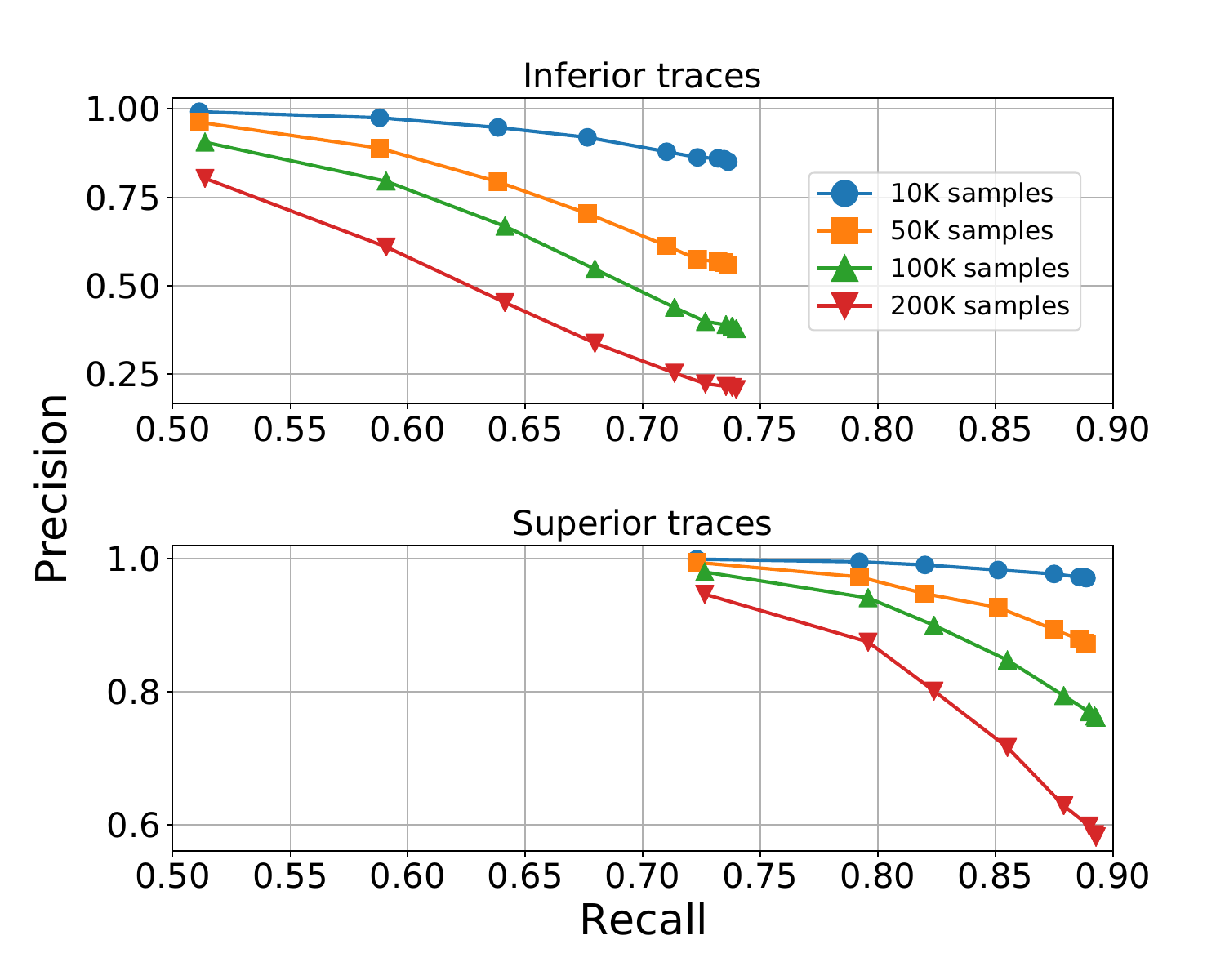}
  \caption{Precision and recall of \ourWF over \lowbw and \highbw traces with larger open world dataset. }
  \label{fig:ow-diff-size}
\end{figure}

For the previous open-world experiments, we used 10K unmonitored website traces. However, there are millions of active websites on the Internet. We evaluate \ourWF with different numbers of unmonitored website traces when the adversary is performing the attack. Figure~\ref{fig:ow-diff-size} shows the precision-recall curve of \ourWF using AWF-OW10k, AWF-OW50k, AWF-OW100k, and AWF-OW200k unmonitored traces. 
The results show that the performance of \ourWF decreases with increasing the open world size for both \lowbw and \highbw traces, e.g., with 200K unmonitored traces, \ourWF has only $25\%$ precision while having $74\%$ recall on \lowbw traces. For 50K unmonitored traces, \ourWF can still achieve $55\%$ precision while maintaining the $74\%$ recall. 
For \highbw traces, even with 100K unmonitored traces, \ourWF has relatively high precision, $75\%$, with $90\%$ recall. 

\begin{figure}[t]
  \centering
  \includegraphics[width = 0.9\linewidth]{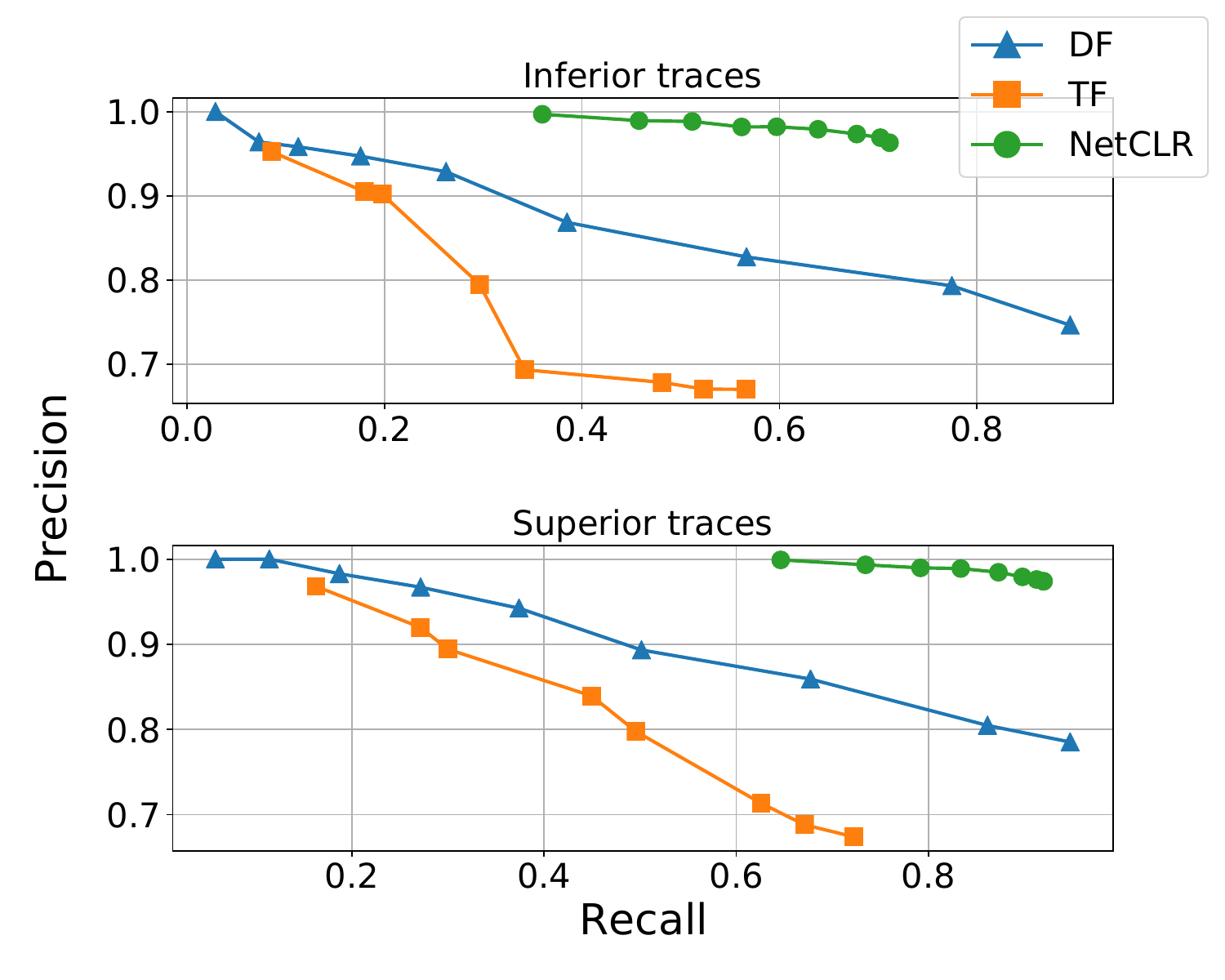}
  \caption{Comparing precision and recall of \ourWF with DF and TF over \lowbw and \highbw traces in the presence of concept drift. \ourWF has the highest $F_1$ score.}
  \label{fig:ow-comp-CD}
\end{figure}

\paragraphb{\Drf Effect:}
We also perform the same experiment when the distribution of pre-training data is different from the distribution of fine-tuning and test data to evaluate the performance of \ourWF against \drf. 
Table~\ref{fig:ow-comp-CD} compares the precision-recall curve of \ourWF with DF and TF.
We used 1000 \highbw unmonitored traces randomly picked from \driftow to perform fine-tuning. For evaluation, we use 4000 of each \lowbw and \highbw unmonitored traces from \driftow.
We use \driftcw as the monitored traces. 
\ourWF has higher precision compared to other attacks. Compared to AWF-OW, \ourWF has lower recall which is expected due to \drf effect. DF still has a higher recall for lower thresholds indicating its ability to distinguish monitored websites from unmonitored ones. However, \ourWF outperforms both DF and TF when comparing $F1$ score indicating that \aug makes \ourWF more robust against \drf.
\textbf{The overall results suggest that \ourWF is more resilient against \drf in an open-world scenario compared to previous attacks while the model is evaluated on traces in unobserved settings.}

\section{Ablation Study}
In this section, we perform an ablation study to better understand why \ourWF is able to outperform SOTA. Due to the number of experiments in our ablation study, we focus on AWF dataset and $N=10$ for all the hyper-parameters. We perform an ablation study on 6 hyper-parameters. 5 of these hyper-parameters belong to \aug and the pre-training phase of \ourWF. We also study the effect of $learning\_rate$ of the fine-tuning phase of \ourWF. 
Table~\ref{tab:abl} shows the accuracy of \ourWF on both \lowbw and \highbw traces with different hyper-parameters. As can be seen, different configurations of hyperparameters for \aug do not cause a great deviation in the accuracy of \ourWF. Table~\ref{tab:abl} shows that \ourWF's performance is not hugely sensitive to the \aug hyper-parameters. For each hyper-parameter, we pick the value with the best accuracy on \lowbw traces as this is the main purpose of this work: A WF attack that is able to achieve high accuracy on unobserved traces. 

Different values for $learning\_rate$, however, can affect the performance of \ourWF drastically, e.g., a learning rate of $10^{-5}$ reduces the accuracy of \ourWF to $77.4\%$ on \lowbw traces which is significantly smaller than the $86.1\%$ accuracy achieved by a learning rate of $5\times10^{-4}$.

\begin{table}
    \centering
    \caption{Accuracy of \ourWF with different hyper-parameters of \aug and $learning\_rate$ for ablation study. We perform the attack with 10 labeled samples. All numbers are $\%$.}
    \resizebox{0.7\columnwidth}{!}{
    \begin{tabular}{@{}l|l|ll@{}}
    \toprule
     & & Accuracy &  \\
    \cmidrule{3-4}
    Hyper-parameter & Value & \Lowbw & \Highbw \\
    \cmidrule(r){1-2}\cmidrule(lr){3-4}
    $n_{merge}$ & 3 & $84.7$ & $94.9$ \\
     & 4 & $83.8 \pm 2$ & $94.7$ \\
     & \textbf{5} & $\bm{86.1 \pm 1.2}$ & $\bm{94.8}$ \\
     & 6 & $84.9 \pm 1.2$ & $94.7$ \\
    \cmidrule(r){1-2}\cmidrule(lr){3-4}
    $r_{merge}$ & 0.05 & $83.4 \pm 1.3$ & $94.3$ \\
    & \textbf{0.1} & $\bm{86.1 \pm 1.2}$ & $\bm{94.8}$ \\
    & 0.2 & $83.7 \pm 1.7$ & $94.5$ \\
    & 0.3 & $83.1$ & $94.4$ \\
    \cmidrule(r){1-2}\cmidrule(lr){3-4}
    $r_{insert}$ & 0.1 & $84.8$ & $94.8$ \\
    & \textbf{0.3} & $\bm{86.1 \pm 1.2}$ & $\bm{94.8}$ \\
    & 0.5 & $83.7 \pm 2.3$ & $94.6$ \\
    & 0.7 & $84.8 \pm 0.9$ & $94.5$ \\
    \cmidrule(r){1-2}\cmidrule(lr){3-4}
    $SHIFT$ & 5 & $84 \pm 2.1$ & $94.8$ \\
    & \textbf{10} & $\bm{86.1 \pm 1.2}$ & $\bm{94.8}$ \\
    & 20 & $84.2 \pm 1.2$ & $94.7$ \\
    & 50 & $81.7 \pm 1.4$ & $93.8 \pm 1$ \\
    \cmidrule(r){1-2}\cmidrule(lr){3-4}
    \textit{embedding\_dim} & 128 & $83.6 \pm 1.3$ & $94.8$ \\
    & 256 & $83.6 \pm 1.3$ & $94.7$ \\
    & \textbf{512} & $\bm{86.1 \pm 1.2}$ & $\bm{94.8}$ \\
    & 1024 & $83.6 \pm 1.3$ & $94.7$ \\
    \cmidrule(r){1-2}\cmidrule(lr){3-4}
    \textit{learning\_rate} & $10^{-5}$ & $77.4 \pm 0.9$ & $89.9$ \\
    & $5\times10^{-5}$ & $83.7$ & $93.9$ \\
    & $10^{-4}$ & $85.3$ & $94.5$ \\
    & $\bm{5\times10^{-4}}$ & $\bm{86.1 \pm 1.2}$ & $\bm{94.8}$ \\
    & $10^{-3}$ & $85$ & $94.9$ \\
    & $5\times10^{-3}$ & $74 \pm 1.3$ & $90.5$ \\
    & $10^{-2}$ & $72.2 \pm 3.3$ & $85.7 \pm 2.4$ \\
    \bottomrule
    \end{tabular}
    }
    \label{tab:abl}
\end{table}

\section{Discussion of Countermeasures}\label{sec:def}
In this section, we assess the performance of NetCLR against one of the leading WF countermeasure techniques, Blind Adversarial Perturbations (BAP), proposed by Nasr et al.~\cite{nasr2021defeating}. BAP is a generic approach that applies adversarial perturbations on live Tor traffic.
BAP trains a neural network that is able to generate adversarial perturbations independent of the incoming Tor trace. 
BAP is also implemented as a Tor pluggable transport. 
In~\cite{nasr2021defeating}, the authors assume that the defender has access to a subset of the training data the original WF classifier is trained on. 
Furthermore, they show that BAP is transferable and is effective in both white-box and black-box scenarios. 
Tik-Tok~\cite{rahman2020tik} is also another SOTA defense mechanism against WF attacks. Tik-Tok focuses on the performance gains obtained by combining timing and direction information in what they call directional timing. However, the results in Tik-Tok paper show that for undefended traces which is the same scenario as our attack model, there is no performance gain in using directional timing compared to only using directions in DF. 
Furthermore, Tik-Tok compares using slow and fast circuits as their test set. The authors show that when training undefended traces, the performance gains from using directional timing are insignificant. We also found no evidence that Tik-Tok which focuses on DF outperforms TF which is a limited-data N-shot learning technique.
Hence, we only focused on evaluating \ourWF against BAP defense mechanism.

To evaluate \ourWF against BAP, we assume a stronger defender with white-box access to the base-model of \ourWF. 
We also assume that for all the values of $N$, the defender has access to all the labeled samples giving them the ability to learn more effective adversarial perturbations. 
Nasr et al. proposed different methods to perturb Tor traces, e.g.,  adding network jitter to inter-packet delays, inserting dummy packets to modify the sizes of  packets, and injecting adversarial directions. BAP by injecting adversarial perturbations is proven to be highly effective in degrading the accuracy of WF classifiers, e.g., they show that using only $2\%$ bandwidth overhead, BAP can reduce the accuracy of DF classifier by $49\%$ which is higher than other countermeasures. 
Since \ourWF uses cell directions as input representations, we use BAP to inject adversarial directions into the traces. 
The bandwidth overhead of this method is defined by a parameter $\alpha$ which represents the number of adversarial directions BAP injects into the trace. 

For this experiment, we consider a closed-world scenario. We use the same pre-trained base model as the previous experiments. We use AWF-{attack} for both fine-tuning and evaluation.
Table~\ref{tab:def-ours} shows the performance of \ourWF on both \lowbw and \highbw traces of AWF-{attack} and for different values of $N$ when the defender injects adversarial directions. 
We observe that even with $2\%$ bandwidth overhead ($\alpha = 100$), \ourWF still has $70\%$ accuracy on \lowbw traces when $N = 10$. This is significantly higher than the performance of DF against BAP, e.g., when $N = 10$ and with $2\%$ bandwidth overhead, the accuracy of DF on \lowbw traces reduces to $12.6\%$. Table~\ref{tab:def-df} in Appendix~\ref{app:def} shows the performance of DF against BAP with different values of $N$ and bandwidth overheads. 
These results show that \textbf{\ourWF is more robust against countermeasure techniques which are due to the pre-training phase of \ourWF and the tailored augmentations that help the model perform better when faced with unobserved traces.} 

\begin{table}
    \centering
    \caption{Accuracy of \ourWF against BAP defense technique with different bandwidth overheads over \lowbw and \highbw traces. As opposed to DF, injecting adversarial directions do not significantly reduce the performance of \ourWF. All numbers are $\%$.}
    \resizebox{\columnwidth}{!}{
    \begin{tabular}{@{}l|ll|ll|ll@{}}
        \toprule
         & No Defense && $\alpha = 50$ && $\alpha = 100$ \\
        \cmidrule{2-7}
        N & \lowbw & \highbw & \lowbw & \highbw & \lowbw & \highbw \\
        \cmidrule(r){1-1}\cmidrule(lr){2-3}\cmidrule(lr){4-5}\cmidrule{6-7}
        5 & $80$ & $92.1$ & $73.5$ & $86.8$ & $56.6$ & $70$ \\
        10 & $84.5$ & $94.2$ & $80.3$ & $91.2$ & $70.4$ & $82.7$\\
        20 & $88.4$ & $96.1$ & $83.7$ & $93.8$ & $66.9$ & $80.1$ \\
        90 & $93.6$ & $97.9$ & $89.2$ & $96.9$ & $71.7$ & $83.7$ \\
        \bottomrule
    \end{tabular}
    }
    \label{tab:def-ours}
\end{table}

\section{conclusion}
In this work, we propose that one of the major limitations of existing website fingerprinting (WF) techniques is their lack of longitudinal perspective into network traffic when training the classifier. 
To alleviate this problem, we propose the use of data augmentation as a potential solution. Specifically, we introduce \aug, an augmentation technique specifically designed for Tor traces, enabling the WF model to classify traces in unobserved settings.
We instantiate \aug through \semi and \self to reduce the reliability of WF attacks on labeled data. 
We then propose \ourWF, a WF attack based on \self and \aug.
Through extensive experiments in both closed-world and open-world scenarios, we demonstrate that \ourWF outperforms existing WF techniques in a realistic scenario where the model is trained on traces from one setting and evaluated on traces from a different setting. Our experiments also show that \ourWF is more resilient against \drf in this realistic scenario.  

\begin{acks}
This work was supported in part by the NSF grant CNS-1953786,
and by the Young Faculty Award program of the Defense Advanced Research Projects Agency (DARPA) under the grant DARPA-RA-21-03-09-YFA9-FP-003.
The views, opinions, and/or findings expressed are those of the authors and should not be interpreted as representing the official views or policies of the Department of Defense or the U.S. Government.
\end{acks}


\printbibliography


\appendix

\section{Semi-Supervised Learning Formulation}\label{app:semi}
In this section, we overview the detailed formulation of FixMatch~\cite{sohn2020fixmatch}.
For an $L$-class classification problem, let $\mathcal{X} = \big\{ (x_b,p_b): b \in (1,\cdots,B) \big\}$ be a batch of $B$ labeled samples, where $x_b$ are the training samples and $p_b$ are one-hot labels. 
Let $\mathcal{U} = \big\{  u_b: b \in (1, \cdots, \mu B) \big\}$ be a batch of $\mu B$ unlabeled samples where $\mu$ is a hyperparameter denoting the ratio of the number of unlabeled samples ($\mathcal{U}$) to labeled samples ($\mathcal{X}$). 
We assume $p_m (y|x)$ is the predicted class distribution produced by the underlying model for input $x$ and $H(p,q)$ denotes the cross-entropy between two probability distributions $p$ and $q$. 
FixMatch performs two types of augmentations: strong and weak, denoted by $\mathcal{A}(.)$ and $\alpha(.)$ respectively. 
The FixMatch loss function consists of two cross-entropy loss terms: a supervised loss $\ell_s$ applied to labeled data and an unsupervised loss $\ell_u$. 
In particular, $\ell_s$ can be the standard cross-entropy loss used in traffic analysis classification scenarios (i.e., WF) on weakly augmented labeled samples:

\begin{align}
    \ell_s = \frac{1}{B} \sum_{b=1}^{B} \text{H} \big( p_b,p_m(y | \alpha(x_b)) \big)
\end{align}

For each unlabeled sample, FixMatch generates a pseudo label which is then used in a standard cross-entropy loss. Given a weakly-augmented version of an unlabeled sample $q_b = p_m (y | \alpha(u_b))$, we compute the model's predicted class distribution. We then use $\hat{q}_b = \arg\max(q_b)$ as a pseudo-label, except we apply the cross-entropy loss against the model's output for a strongly-augmented version of $u_b$:

\begin{align}
    \ell_u = \frac{1}{\mu B} \sum_{b=1}^{\mu B} \mathbb{1}\big(\max(q_b) \geq \tau_f \big) \text{H}\big( \hat{q}_b, p_m (y | \mathcal{A}(u_b)) \big)  
\end{align}

\noindent where $\tau_{f}$ is a scalar hyperparameter indicating the threshold above which we retain a pseudo-label. We minimize the following loss: $\ell_s + \lambda_u \ell_u$ where $\lambda_u$ is a fixed scalar hyperparameter denoting the relative weight of the unlabeled loss.

\begin{figure*}[!t]
  \centering
  \includegraphics[width = 0.8\linewidth]{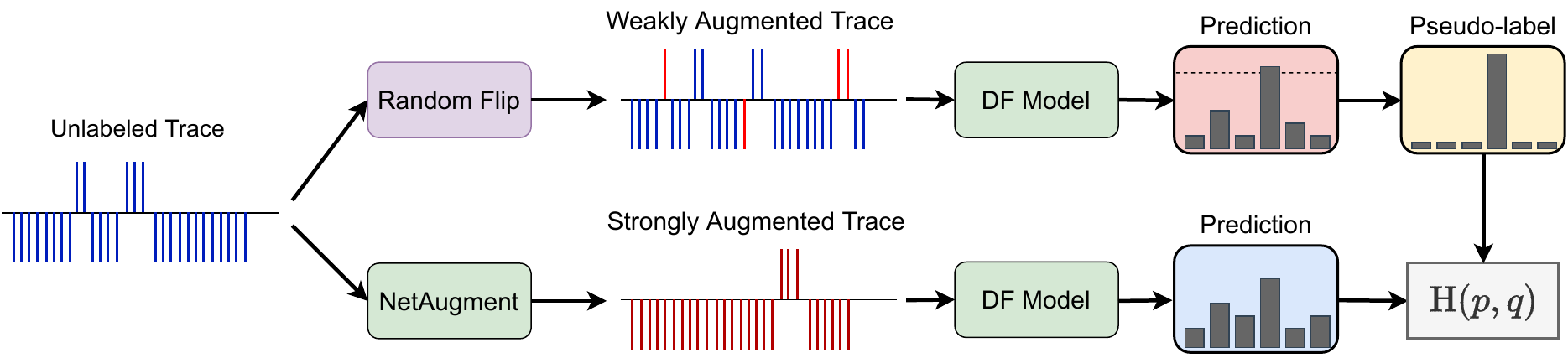}
  \caption{Diagram of \netfm when fed with network traces}
  \label{fig:fixmatch}
\end{figure*}

\section{Self-Supervised Learning Formulation}\label{app:self}
SimCLR framework comprises the following four major components:
\begin{compactitem}
    \item A stochastic data augmentation module that transforms any given input sample $x$ resulting in two correlated views of the same sample, denoted $\tilde{x}_i$ and $\tilde{x}_j$, which we consider as a positive pair. 
    \item A neural network base encoder $f(.)$ that extracts embeddings from augmented samples. We use the underlying neural works used in traffic analysis algorithms to obtain $e_i = f(\tilde{x}_i)$ where $e_i \in \mathbb{R}^d$ is the output of the underlying neural network before the last layer (The fully connected layer that generates the output probabilities); $e_i$ also represents the sample embedding extracted by the base model. 
    \item A small neural network \textit{projection head} that maps embeddings to the space where contrastive loss is applied. The projection head is a fully connected network with one hidden layer to obtain $z_i = g(e_i) = W^{(2)}\sigma (W^{(1)}e_i)$ where $\sigma$ is a ReLU non-linearity. 
    \item A \textit{contrastive loss} function to maximize the agreement between augmented versions of an unlabeled sample. Given a set $\{ \tilde{x}_k \}$ including a positive pair of examples $\tilde{x}_i$ and $\tilde{x}_j$, the contrastive prediction task aims to identify $\tilde{x}_j$ in $\{ \tilde{x}_k \}_{k \neq i}$ for a given $\tilde{x}_i$.  
\end{compactitem}

SimCLR randomly samples a minibatch of $N$ samples and applies contrastive loss on pairs of augmented samples generated from the minibatch, resulting in 2$N$ network traces. To construct the negative samples, given a positive pair, SimCLR considers the other 2$(N-1)$ augmented samples within a minibatch as negative samples. If we assume $\text{sim}(u, v) = u^T v / \lVert u \rVert \lVert v \rVert$ denotes the dot product between $\ell_2$ normalized vectors of $u$ and $v$ (i.e. cosine similarity), the contrastive loss function for a positive pair of samples $(i, j)$ is defined as:

\begin{align}\label{eq:contrasive-loss}
    \ell_{i,j} = - \log \frac{\exp(\text{sim}(z_i, z_j)/\tau_s)}{\sum_{k=1}^{2N} \mathbb{1}_{[k\neq i]}\exp (\text{sim}(z_i, z_k)/\tau_s)}
\end{align}

\noindent where $\mathbb{1}_{[k\neq i]} \in \{0,1\}$ is an indicator function evaluating to 1 iff $k \neq i$ and $\tau_s$ denotes a temperature parameter. The SimCLR work term this loss \textit{NT-Xent} (the normalized temperature-scaled entropy loss).


\section{Analyzing Concept Drift}\label{app:drift}
In this section, we analyze the actual observed concept drift between \driftcw and AWF-{attack} datasets. In particular, we calculate the degradation in accuracy caused by concept drift. Table~\ref{tab:drift-analyze} shows the results. As the results suggest, \ourWF caused less decrease in the accuracy indicating that \ourWF is more robust against concept drift.

\begin{table*}
    \centering
    \caption{Difference between the accuracy of AWF-{attack} dataset and \driftcw dataset. \ourWF is more robust against concept drift. All numbers are $\%$.}
    \resizebox{\textwidth}{!}{
    \begin{tabular}{@{}l|lll|lll|lll@{}}
        \toprule
         & DF~\cite{sirinam2018deep} & & & \ourWF \footnotesize{(\flipaug)}  & & & \ourWF \footnotesize{(\aug)} \\
         \cmidrule{2-10}
        N & AWF-{attack} & \driftcw & Difference & AWF-{attack} & \driftcw & Difference & AWF-{attack} & \driftcw & Difference \\
         \cmidrule(r){1-1}\cmidrule(lr){2-4}\cmidrule(lr){5-7}\cmidrule(lr){8-10}
        5 &  $60.9$ & $49$ & $11.9$ & $80.7$ & $69.7$ & $11$ & $89.7$ & $79.3$ & $10.4$ \\
        
        10 &  $78.1$ & $70.3$ & $7.8$ & $90.5$ & $82.4$ & $6.1$ & $94.5$ & $89.2$ & $5.3$\\
        
        20 &  $86.1$ & $82.1$ & $4$ & $94.4$ & $89.2$ & $5.2$ & $96.6$ & $93.6$ & $3$ \\
        \bottomrule
    \end{tabular}
    }
    \label{tab:drift-analyze}
\end{table*}

\section{Network Condition Metric}
\begin{figure}[!t]
  \centering
  \resizebox{\columnwidth}{!}{
  \includegraphics[height = 0.8\textheight,width=\textwidth,keepaspectratio]{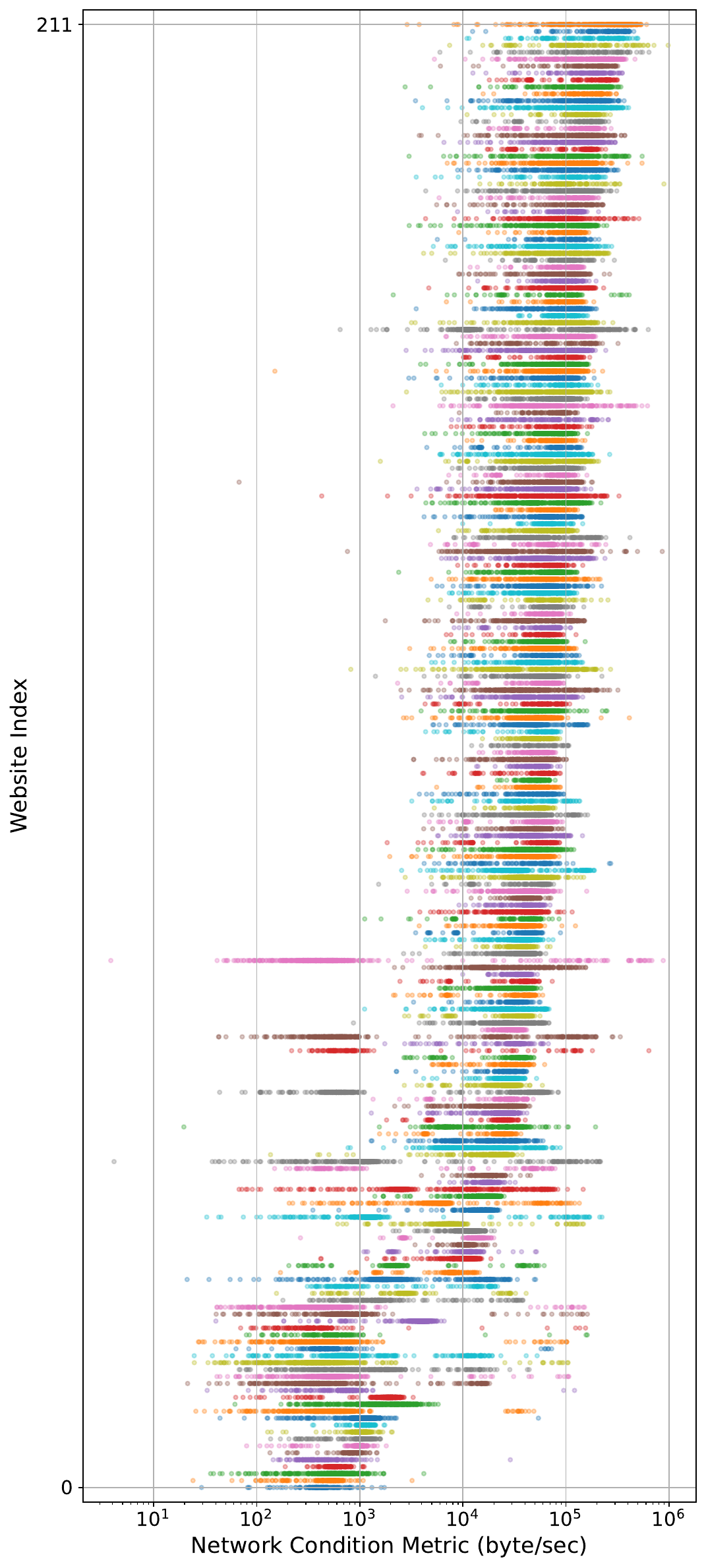}
  }
  \caption{Diagram of NCM values for traces of websites in the Drift dataset}
  \label{fig:ncm-drift}
\end{figure}

Figure \ref{fig:ncm-drift} show the NCM values for traces of 211 of the non-onion websites visited to create the Drift dataset. The markers in each row represent the NCM value for different traces of the same website. 

\section{BAP Results}\label{app:def}

Table~\ref{tab:def-df} shows the performance of DF against the BAP countermeasure using different numbers of labeled samples for fine-tuning. As we mentioned in Section~\ref{sec:def}, \ourWF is significantly more robust against BAP. 

\begin{table*}[!t]
    \centering
    \caption{Accuracy of DF~\cite{sirinam2018deep} against BAP defense technique with different bandwidth overheads over \lowbw and \highbw traces. Injecting adversarial directions reduces the performance of DF significantly.}
    \begin{tabular}{@{}l|ll|ll|ll@{}}
        \toprule
         & No Defense && $\alpha = 50$ && $\alpha = 100$ \\
        \cmidrule{2-7}
        N & \lowbw(\%) & \highbw(\%) & \lowbw(\%) & \highbw(\%) & \lowbw(\%) & \highbw(\%) \\
        \cmidrule(r){1-1}\cmidrule(lr){2-3}\cmidrule(lr){4-5}\cmidrule{6-7}
        5 & $48.8$ & $55.3$ & $34.4$ & $37.6$ & $21.4$ & $23.5$\\
        10 & $62.9$ & $78.9$ & $30$ & $40$ & $12.6$ & $15.2$\\
        20 & $76.9$ & $89.4$ & $26$ & $31.6$ & $10.9$ & $15.5$\\
        90 & $84.3$ & $96.1$ & $37.4$ & $49$ & $15.8$ & $19$\\
        \bottomrule
    \end{tabular}
    \label{tab:def-df}
\end{table*}

\begin{table*}[t]
    \centering
    \caption{Hyperparameters of \ourWF and \netfm. The hyperparameters for DF are the same as the original paper~\cite{sirinam2018deep}.}
    \resizebox{0.7\textwidth}{!}{
    \begin{tabular}{||c|c|c|c||}
        \hline
         & \textbf{\ourWF Pre-Training} & \textbf{\ourWF Fine-Tuning} & \textbf{\netfm} \\
        \hline
        Learning Rate & $3\times10^{-4}$& $5\times10^{-4}$ & $10^{-2}$\\
        \hline
        Optimizer & Adam with Cosine Scheduler & Adam & SGD (momentum = 0.9)\\
        \hline
        Input Size & 5000 & 5000 & 5000\\
        \hline
        embedding\_dim & 512 & --- & --- \\
        \hline
        Output Size & embedding\_dim/4 & Number of Classes & Number of Classes\\
        \hline
        Batch Size & 256 & 32 & 256\\
        \hline
        Epochs & 100 & 30 & 100\\
        \hline
        $\lambda_u$ & --- & --- & 1 \\
        \hline
        $\tau$ & $\tau_f = 1$ & --- & $\tau_s = 1$ \\
        \hline
        $p_{flip}$ & 0.1 & --- & 0.1 \\
        \hline
    \end{tabular}
    }
    \label{tab:param}
\end{table*}

\section{Concept Drift in WF Attacks}\label{app:drf}
In a dynamic environment, the distribution of the target variable that a machine learning model learns can be non-stationary and therefore change over time, impacting the performance of the model when deployed. In the literature, this change in the distribution is referred to as dataset shift~\cite{morenotorres2012drift} or concept drift~\cite{gama2014conceptsurvey}.
In the classification context, a problem can be defined by a set of covariates $X$, denoting the features, a class (target) variable $y$, and a joint distribution $P(y,X)$ referred to as concept, which can be written as $P(X|y)P(y)$. Therefore, concept drift between time $t_0$ and time $t_1$ can be defined as $\exists X: P_{t_0}(X,y) \neq P_{t_1}(X,y)$, where $P_{t_i}(X,y)$ denotes the joint distribution between X and y at time $t_i$ for $i \in \{0,1\}$. \textit{Real concept drift} is the changes in $P(X|y)$ between $t_0$ and $t_1$ which can happen whether $P(y)$ changes or not. \textit{Virtual concept drift}, also referred to as \textit{covariate shift} and \textit{feature change} happens if $P(X)$ changes between $t_0$ and $t_1$ but $P(y|X)$ does not change.

The data we use for our experiments consists AWF, a large dataset collected in 2017~\cite{rimmer2017automated}, as well a smaller Drift dataset that we collect 5 years later.
Over time, the content of a website changes and can be completely different after 5 years, e.g. its back-end can be different resulting in a different loading time. Yet, it still has the same class label (URL), meaning $P(X|y)$ changes over time. Therefore, a deployed WF model needs to be resilient to real concept drift.
Note that this requirement exists whether or not the WF model uses offline learning or online learning. Traditional WF attacks ~\cite{oh2021gandalf, sirinam2019triplet, var-cnn, rimmer2017automated} use offline learning where all of the training data is available at the time of model training. Cherubin et al. ~\cite{cherubin2022online} propose using online learning, where the deployed model is continuously updated as more training data arrives. In either case, in a non-stationary and dynamic environment, the data evolves over time as the content of websites changes and new websites are added to the monitored set. Therefore, $P(X|y)$ is changing over time.

Furthermore, our method of collecting traffic and extracting Tor cells for generating traces of a website slightly differs from that of \cite{rimmer2017automated}. For instance, the amount of time we wait for each website to load is different. Consequently, the distribution of traces collected for the same website can vary between the two datasets, even if everything else is the same. In this case, $P(X)$ has changed, but $P(y|X)$ has not changed. This is an instance of virtual concept drift to which our WF model should be resilient.

\section{Model Hyperparameters}\label{app:param}

Table~\ref{tab:param} shows the hyperparameters of \ourWF and \netfm and their optimal values. We select each hyperparameter by searching through a set of candidates. We pick the hyperparameter which leads to the best performance.

\section{Effect of Circuit Bandwidth}\label{app:consensus}

To provide more evidence that \ourWF is resilient to variations in traces under disparate network conditions, we fine-tune the model using \driftcw traces collected through circuits whose guard relay has a high consensus bandwidth and test the model on traces collected through gateways whose guard relay has a low consensus bandwidths. As described in Section \ref{sub:driftdataset}, we logged the bandwidth file information while collecting traces for the Drift dataset. The bandwidth file includes the consensus bandwidth of Tor relays in bytes per second. We found 25 MBps to be the appropriate threshold to split the traces based on the consensus bandwidth of their guard relay as the performance of SOTA WF attacks drops with a lower threshold.
We 
generate training, validation, and test sets with an equal number of traces from each bandwidth category such that there are 20 samples per website from each category in each of the validation and test sets. Table~\ref{tab:consensus} shows the performance of \ourWF when trained using $N=\{5, 10, 20, 90\}$ labeled samples from traces in the high-bandwidth category. \ourWF outperforms both DF and TF when dealing with both \drf and guard relays with low consensus bandwidth.

\begin{table}
    \centering
    \caption{Comparing the accuracy of \ourWF with DF and TF when fine-tuning and testing traces are collected using circuits with different consensus bandwidths. \ourWF outperforms other models when faced with traces in low-bandwidth circuits. All numbers are $\%$. We do not show standard deviations less than $1\%$.}
    \resizebox{\columnwidth}{!}{
    \begin{tabular}{@{}l|ll|ll|ll@{}}
        \toprule
         & DF~\cite{sirinam2018deep} & & TF~\cite{sirinam2019triplet} & & \ourWF \\
         \cmidrule{2-7}
        N & Low BW & High BW & Low BW & High BW & Low BW & High BW \\
         \cmidrule(r){1-1}\cmidrule(lr){2-3}\cmidrule(lr){4-5}\cmidrule(lr){6-7}
        5 & $35.1 \pm 1.3$ & $37.8$ & $47.8$ & $52.2$ & $\bm{65.1 \pm 1.2}$ & $\bm{69.9}$  \\
        10 & $47 \pm 1.1$ & $50.9$ & $54.2$ & $59.4$ & $\bm{76.4 \pm 1}$ & $\bm{80.5}$ \\
        20 & $55.7 \pm 1.3$ & $61.1 \pm 1.3$ & $59.2$ & $64.3$ & $\bm{82.8}$ & $\bm{85.9}$  \\
        90 & $75.1$ & $79.6$ & $67.7$ & $72.9$ & $\bm{89.8}$ & $\bm{92.5}$ \\
        \bottomrule
    \end{tabular}
    }
    \label{tab:consensus}
\end{table}


\section{Training Time Comparison}\label{app:time}
In this Section, we compare the time required to train \ourWF and \dfsame. 
Table~\ref{tab:time} shows the training time for different values of $N$. 
As the results illustrate, \ourWF is two orders of magnitude faster than \dfsame. This is because \ourWF is pre-trained only once and the adversary can use the pre-trained model to fine-tune on the small target dataset. 
On the other hand, the adversary needs to train \dfsame on the whole pre-training and fine-tuning dataset which takes a much longer time.

\begin{table}
    \centering
    \caption{Comparing training time of \ourWF and \dfsame. \ourWF is faster to train by two orders of magnitude. }
    \resizebox{0.5\columnwidth}{!}{
    \begin{tabular}{@{}l|ll@{}}
        \toprule
         & Training Time & \\
        \cmidrule{2-3}
        N & \dfsame & \ourWF \\
        \midrule
        5 & 465.4 s & 3.3 s \\
        10 & 469.4 s & 6.5 s \\
        20 & 518.31 s & 13.7 s \\
        90 & 553.63 s & 60.6 s \\
        \bottomrule
    \end{tabular}
    }
    \label{tab:time}
\end{table}


\clearpage

\end{document}